\documentclass[12pt]{article}
\usepackage{amsmath, amsfonts, amscd, amssymb}
\usepackage{amsthm}
\newcommand{\aconn}{\mathcal{A}}

\newcommand{\bra}[1]{\pmb{\langle}#1\pmb{|}}
\newcommand{\braket}[2]{{\langle #1 |\, #2\rangle}}

\newcommand{\conn}{\mathcal{D}}

\newcommand{\curv}{\mathcal{F}}

\newcommand{\diff}{\mathbf{\Omega}}

\newcommand{\gauge}{\mathcal{U}}
\newcommand{\kd}{\text{\texttt{d}}}
\newcommand{\bkd}{\text{\large\texttt{d}}}

\newcommand{\ket}[1]{\pmb{|}#1\pmb{\rangle}}

\newcommand{\modl}{\mathbf{\mathcal{E}}}

\newcommand{\cont}{\mathcal{C}^{0}}
\newcommand{\smooth}{\mathcal{C}^{\infty}}
\newcommand{\spn}{{\rm span}}

\newcommand{\com}{\mathbb{C}}
\newcommand{\A}{\mathbb{A}}
\newcommand{\mapto}{\longrightarrow}

\newcommand{\ketbra}[2]{\ket{#1}\bra{#2}}

\newcommand{\N}{\mathbb{N}}
\newcommand{\Z}{\mathbb{Z}}
\newcommand{\R}{\mathbb{R}}

\title{\bf Finitary \v{C}ech-de Rham Cohomology:\\
much ado without $\smooth$-smoothness}

\author{A. Mallios\thanks{Algebra and Geometry Section,
Department of Mathematics, University of Athens,
Panepistimioupolis 157 84, Athens, Greece; e-mail:
amallios@cc.uoa.gr} and I. Raptis\thanks{Theoretical Physics
Group, Blackett Laboratory of Physics, Imperial College of
Science, Technology and Medicine, Prince Consort Road, South
Kensington, London SW7 2BZ, UK; e-mail: i.raptis@ic.ac.uk}}

\date{}

\begin{document}

\maketitle

\begin{abstract}

\noindent The present paper continues \cite{malrap} and studies
the curved finitary spacetime sheaves of incidence algebras
presented therein from a \v{C}ech cohomological perspective. In
particular, we entertain the possibility of constructing a
non-trivial de Rham complex on these finite dimensional algebra
sheaves along the lines of the first author's axiomatic approach
to differential geometry via the theory of vector and algebra
sheaves \cite{mall1, mall2}. The upshot of this study is that
important `classical' differential geometric constructions and
results usually thought of as being intimately associated with
$\smooth$-smooth manifolds carry through, virtually unaltered, to
the finitary-algebraic regime with the help of some quite
universal, because abstract, ideas taken mainly from
sheaf-cohomology as developed in \cite{mall1, mall2}. At the end
of the paper, and due to the fact that the incidence algebras
involved have been interpreted as quantum causal sets \cite{rap1,
malrap}, we discuss how these ideas may be used in certain aspects
of current research on discrete Lorentzian quantum gravity.

\end{abstract}

\section{The general question motivating our quest}\label{sec1}

\begin{itemize}

\item How much from the differential geometric panoply of
$\smooth$-smooth manifolds can we carry through, almost intact, to
a finitary ({\it ie}, locally finite) algebraic setting?

\end{itemize}

\noindent This is the general question that motivates the present
study. We will also ponder on the following question that is
closely related to the one above, but we will have to postpone our
detailed elaborations about it for a future work, namely:

\begin{itemize}

\item Are the pathologies ({\it eg}, the so-called singularities)
of the usual differential calculus on smooth manifolds `innate' to
the calculus or `differential mechanism' itself, or are they due
to the particular structure (commutative algebra) sheaf of the
infinitely differentiable functions that {\em we} employ to
coordinatize the points of the $\smooth$-smooth manifold?

\end{itemize}

\noindent The latter question, which in our opinion is the deeper
of the two, puts into perspective the classical diseases in the
form of infinities that assail both the classical and the quantum
field theories of the dynamics of spacetime ({\it ie}, gravity)
and matter ({\it ie}, gauge theories), which theories, in turn,
assume up-front a smooth base spacetime continuum on which the
relevant smooth fields are localized, dynamically propagate and
interact with each other. For if these pathologies ultimately turn
out to be not due to the differential mechanism itself, but rather
due to our own assumption of algebras of $\smooth$-smooth
coordinatizations (or measurements!) of the manifold's point
events, there is certainly hope that by changing focus from the
structure sheaf of rings of infinitely differentiable functions on
the smooth manifold to some other `more appropriate' (or {\em
suitable to the particular physical problem in focus}) algebra
sheaves, while at the same time retaining at our disposal most (in
effect, {\em all}!) of the powerful differential geometric
constructions and techniques, the aforementioned diseases may be
bypassed or even incorporated into the resulting `generalized and
abstract differential calculus' \cite{mall1, mall2}, something
that would effectively indicate that they are not really an
essential part of `the problem' after all \cite{mall3,
mall4}---that is, if there still is any problem left for us to
confront\footnote{Certainly, there will still remain the noble
challenge to actually construct a conceptually sound and
`calculationally' finite quantum theory of gravity, but at least
it will have become clear that the singularities of classical
gravity and the weaker but still stubborn infinities of quantum
field theory are due to an inappropriate assumption---that of
$\smooth$-smooth coordinates, not a faulty mechanism---that of the
differential calculus, and as a result they should present no
essential, let alone insuperable, obstacles on our by now
notoriously long (mainly due to these pathologies of the
$\smooth$-smooth manifold) way towards the formulation of a cogent
quantum gravity. For instance, the works \cite{malros1, malros2}
nicely capture this spirit, namely, that one can actually carry
out the usual differential geometric contsructions over spaces and
their coordinate structure algebra sheaves that are very singular
and anomalous---especially when viewed from the perspective of the
featureless $\smooth$-smooth continuum.}.

So, to recapitulate our attitude towards the opening two
questions: we contend that the usual differential geometry of the
`classical' $\smooth$-manifolds could be put into an entirely
`algebraic' ({\it ie}, sheaf-theoretic) framework, thus avoid
making use of any Calculus at all, at least in the classical sense
of the latter term. Thus, to a great extent, differentiability may
prove to be, in a deep sense, independent of smoothness and, as a
result, gravity may be transcribed to a reticular-algebraic and
sheaf-theoretic environment more suitable for infusing quantum
ideas into it than the problematic classical geometric
$\smooth$-smooth spacetime continuum. As a bonus from this
transcription, we may discover that in the new finitary setting
the classical smooth differential pathologies are evaded, perhaps
even incorporated into the more general, abstract and of a strong
algebraic character sheaf-theoretic differential geometric picture
\cite{mall1, mall2}, so that they are not essentially contributing
factors to the difficulty of the problem of arriving at a sound
quantum theory of gravity \cite{mall3, mall4}. However, it may
well turn out that in the particular finitary-algebraic sheaf
theoresis of spacetime structure and dynamics favored here, the
real difficulties lie elsewhere, and that they are even more
severe than the ones troubling their smooth counterparts.
Undoubtedly we must keep an open mind, but then again we must also
keep an optimistic eye and, at this early stage of the development
of the theory, at least give such alternative
combinatory-algebraic sheaf-theoretic ideas a decent chance.

We must also admit that such an endeavor is by no means new.
Indeed, Einstein, as early as one year after he presented the
general theory of relativity, doubted in the light of the quantum
the very geometric smooth spacetime continuum that supported his
classical field theory of gravity:

\begin{quotation}

{\footnotesize ``...you have correctly grasped the drawback that
the continuum brings. If the molecular view of matter is the
correct (appropriate) one; {\it ie}, if a part of the universe is
to be represented by a finite number of points, then the continuum
of the present theory contains too great a manifold of
possibilities. I also believe that this `too great' is responsible
for the fact that our present means of description miscarry with
quantum theory. The problem seems to me how one can formulate
statements about a discontinuum without calling upon a continuum
space-time as an aid; the latter should be banned from theory as a
supplementary construction not justified by the essence of the
problem---{\em a construction which corresponds to nothing real.
But we still lack the mathematical structure
unfortunately}\footnote{Our emphasis.}. How much have I already
plagued myself in this way of the manifold!...}" (1916)
\cite{stachel}

\end{quotation}

\noindent and just one year before his death he criticized the
pathological nature of the geometric spacetime continuum so that,
in view of the atomistic character of Physis that the quantum
revolution brought forth, he prophetically anticipated ``{\itshape
a purely algebraic theory for the description of reality}"
\cite{einst56}, much as follows:

\begin{quotation}

``{\footnotesize ...An algebraic theory of physics is affected
with just the inverted advantages and weaknesses, aside from the
fact that no one has been able to propose a possible logical
schema for such a theory. {\em It would be especially difficult to
derive something like a spatio-temporal quasi-order from such a
schema}\footnote{Again, our emphasis in order to prepare the
reader for our quantum causal elaborations in the sequel.}. I
cannot imagine how the axiomatic framework for such a physics
would appear, and I don't like it when one talks about it in dark
apostrophes. But I hold it entirely possible that the development
will lead there; for it seems that the state of any finite
spatially limited system may be fully characterized by a finite
set of numbers. This seems to speak against a continuum with its
infinitely many degrees of freedom. The objection is not decisive
only because {\em one doesn't know, in the contemporary state of
mathematics, in what way the demand for freedom from singularity
(in the continuum theory) limits the manifold of
solutions}\footnote{Again, our emphasis.}...}" (1954)\footnote{It
is quite remarkable indeed that these ideas of Einstein,
especially his anticipation in the second quotation above of
deriving a spatio-temporal quasi-order from a discrete-algebraic
theoretical schema, foreshadow a modern approach to quantum
gravity pioneered by Sorkin and coworkers coined {\em causal set
theory} \cite{bomb87, sork90a, sork90b, sork95, sork97,
ridesork00, sork01}, as well as its reticular-algebraic `{\em
quantum causal set}' outgrowth \cite{rap1, malrap, rap3}. In these
approaches to quantum gravity it is fundamentally posited that
underlying the spacetime manifold of macroscopic experience there
are (quantum) causal set substrata---partially ordered sets (and
their associated incidence algebras) with their order being
regarded as the discrete and quantum ancestor of the
spatio-temporal quasi order encoded in the lightcones of the
classical relativistic spacetime continuum, which classical causal
order, in turn, `derives' from ({\it ie}, can be thought as a
coarse descendant of) the fundamental causal order of causal sets
and their quantal incidence algebraic relatives. We will return to
the causal set idea as well as to its algebraic and
sheaf-theoretic counterparts in some detail in sections \ref{sec6}
and \ref{sec7}. The second author wishes to thank Rafael Sorkin
for discussing the relevance of the Einstein quotation above to
the problem of quantum gravity---especially to the (quantum)
causal set-theoretic approach to the latter problem.}
\cite{stachel}

\end{quotation}

\noindent and a little bit later he agnostically admitted:

\begin{quotation}

``{\footnotesize ...Your objections regarding the existence of
singularity-free solutions which could represent the field
together with the particles I find most justified. I also share
this doubt. If it should finally turn out to be the case, then I
doubt in general the existence of a rational and physically useful
continuous field theory. But what then? Heine's classical line
comes to mind: `{\sl And a fool waits for the answer}'...}" (1954)
\cite{stachel}

\end{quotation}

So, as noted earlier, here we will content ourselves with trying
to answer to the first question opening this paper and in a later
work we will attempt to swim in the depths of the second
\cite{malrap1}. Below, after we give a `crash' review of the basic
ingredients in the first author's Abstract Differential Geometry
(ADG) theory \cite{mall1, mall2} (section \ref{sec2}), we initiate
a \v{C}ech-type of cohomological treatment of the curved finitary
spacetime sheaves (finsheaves) \cite{rap2} of incidence Rota
algebras representing quantum causal sets (qausets) introduced in
\cite{malrap} (section \ref{sec3}), and then construct the
relevant de Rham complex on them based on an abstract version of
de Rham's theorem {\it \`a la} ADG \cite{mall1, mall2} (section
\ref{sec4}). The possibility of recovering the `classical'
$\smooth$-smooth \v{C}ech-de Rham complex from a net of the
aforementioned finsheaf-cochains above will be entertained in
section \ref{sec5}. Having the finitary complex in hand, we will
discuss the possibility of a finitary sheaf-cohomological
classification of the reticular spin-Lorentzian connection fields
$\aconn_{m}$ dwelling on the gauged ({\it ie}, curved) principal
spin-Lorentzian $\mathcal{G}_{m}$-finsheaves of qausets and their
associated vector (state) finsheaves studied in \cite{malrap} in
much the same way that Maxwell fields on appropriate vector (line)
bundles associated with $\mathcal{G}=U(1)$-principal fiber bundles
were classified, and subsequently `prequantized', along
Selesnick's line bundle axiomatics for the second quantization of
bosonic (photon) fields \cite{sel83}, in \cite{mall1, mall2,
mall5, mall6} (section \ref{sec6}). Arguably, as we contend in the
penultimate part of the paper (section \ref{sec7}), sections of
the vector finsheaves associated with the principal
spin-Lorentzian $\mathcal{G}_{m}$-finsheaves of qausets correspond
to states of `bare' or free graviton-like quanta\footnote{Coined
`causons'---the quanta of causality---in \cite{malrap}.}
\cite{mall1, mall2, mall3, mall4, malrap, mall5,
mall6}\footnote{By the way, we also read from \cite{sel83} that
states of second quantized free fermionic fields can be identified
with sections of Grassmannian vector bundles (correspondingly,
vector sheaves in \cite{mall1, mall2, mall5, mall6}).}, so that in
the present granular-algebraic context the fool in Einstein's
quotation of Heine above will appear to have found the answer that
he was desperately waiting for---and, all the more remarkably, by
evading altogether the $\smooth$-smoothness of the classical
geometric spacetime manifold. All in all, we hold that {\em field
and particle can possibly coexist at last} by going around the
differential manifold spacetime and its pestilential singularities
via discrete-algebraic and sheaf-theoretic means\footnote{To put
it differently, and in contrast to Einstein's mildly pessimistic
premonitions above, to us field theory does not appear to be
inextricably tied to a geometric spacetime continuum: one can
actually do field theory on relatively discrete ({\it ie},
`singular' and `disconnected' from the $\smooth$-smooth
perspective) spaces. We thus seem to abide to the general
philosophy that whenever one encounters a contradiction between
the mathematics (model) and the physics (reality), one should
always change the maths. For {\em Nature cannot be pathological};
it is only that our theoretical models of Her are of limited
applicability and validity \cite{malrap}.}. Section \ref{sec7}
closes with a brief discussion of some possible applications of
such finitary-algebraic models and their quantum causal
interpretation \cite{bomb87, sork90a, sork90b, sork95, sork97,
rap1, malrap, rap3, rap4, sork01} to current research on discrete
Lorentzian quantum gravity, as well as highlighting some
suggestive resemblances between our finitary application of ADG
and the Kock-Lawvere Synthetic Differential Geometry (SDG)
\cite{lav96} . The paper concludes (section \ref{sec8}) with some
physico-philosophical remarks in the spirit of the two motivating
opening questions above, as it were, to close the circle that they
opened.

\section{A brief review of Abstract Differential
Geometry}\label{sec2}

The rather technical elements from Mallios' Abstract (Axiomatic)
Differential Geometry (ADG) to be briefly presented below are
selected from \cite{mall2} which is a concise {\it r\'{e}sum\'{e}}
of the more complete, but also more voluminous, work
\cite{mall1}\footnote{With the physicist in mind, we are not
planning to plough through \cite{mall1} in any detail here. Our
`heuristic' presentation of ADG from \cite{mall2} should suffice
for the `physical level of rigour' assumed to be suitable for the
present `physics oriented' study.}. We itemize our brief review of
ADG into four parts: the basic mathematical objects involved, the
main axioms adopted, the central mathematical technique used and
ADG's core philosophy.

\subsection{About the assumptions: three basic sheaves, three basic objects}

The basic mathematical objects involved in the development of ADG
are sheaves of (complex) vector spaces $V$ ($\com$-vector space
sheaves), of (complex abelian) algebras $A$ ($\com$-algebra
sheaves) and of (differential) modules $E$ over such algebras
($A$-module sheaves). These sheaves are generically symbolized by
$\mathcal{V}$, $\A$\footnote{In both \cite{mall1} and \cite{mall2}
commutative algebra sheaves were denoted by $\aconn$. However, the
same symbol we have already reserved for the spin-Lorentzian
connections involved in \cite{malrap}. Thus, `$\A$' will be used
henceforth to symbolize abelian algebra sheaves.} and $\modl$,
respectively.

The first basic object to be associated with the three kinds of
sheaves above is, of course, the base topological space $X$ over
which the vector space objects V dwelling in (the stalks of)
$\mathcal{V}$, the $\com$-algebras $A$ in $\A$ and the $A$-modules
$E$ in $\modl$, are localized. It is one of the principal
assumptions of ADG that all the three basic sheaves above have as
common base space an {\em arbitrary topological space}---although
this generality and freedom of choosing the `localization space'
$X$ is slightly constrained by assuming that it should be, at
least, {\em paracompact} and {\em Hausdorff}. We will return to
these two `auxiliary assumptions' for $X$ in the next two
sections. For now we note that in what follows $X$ will be usually
omitted from the sheaves above ({\it ie}, we will simply write
$\mathcal{V}$, $\A$ {\it etc}, instead of $\mathcal{V}(X)$,
$\A(X)$ {\it etc}) for typographical economy, unless of course we
wish to comment directly on the attributes of $X$. At this point
we should also mention that in \cite{mall1, mall2} an open
covering $\gauge=\{ U\subseteq X:~ U~{\rm open~ in}~ X\}$ of $X$
such that an $A$-module sheaf $\modl(X)$ splits
locally\footnote{That is, with respect to every $U$ in $\gauge$.}
into a finite $n$-fold Whitney (or direct) sum $A^{n}$ of $A$ with
itself\footnote{One may simply think of $A^{n}$ as a finite
dimensional module over $A$---a module of finite `rank' $n$.} as
$\modl|_{U}=\A^{n}|_{U}\simeq{}^{\com}\mathcal{V}^{n}$\footnote{Where
$\A^{n}|_{U}\simeq{}^{\com}\mathcal{V}^{n}$ denotes the
corresponding $n$-dimensional $\com$-vector sheaf isomorphism. We
also note in this context that a $1$-dimensional vector sheaf
({\it ie}, a vector sheaf of rank  $n=1$) is called a `line sheaf'
in ADG.} is called `a local frame of' or `a coordinatizing open
cover of', or even `a local choice of basis (or gauge!) for
$\modl$'\footnote{Accordingly, every covering set $U$ in $\gauge$
is coined `a local gauge of $\modl$'.}. Thus, quite reasonably,
the local sections of the abelian structure $\com$-algebra sheaf
$\A$ relative to the local frame $\gauge$ carry the geometric
denomination `(local) coordinates', while $\A$ itself is called
`the coefficient' or `$c$-number coordinate sheaf' (of $\modl$).

The second essential object involved in ADG is the so-called {\em
$\mathbf{C}$-algebraized space}, represented by the pair $(X,\A)$;
where $X$ is a topological space and $\A$ a commutative
$\com$-algebra sheaf on it. For completeness, perhaps we should
also include $\mathbf{C}$---the constant sheaf of complex numbers
$\com$ over $X$---into the $\mathbf{C}$-algebraized space, but
again for typographical economy we will omit it in our
elaborations below.

The last basic object involved in ADG is the so-called {\em
differential triad}, represented by the triplet
$(X,\modl(\A),\bkd)$; where $X$ is again the base topological
space, $\A$ again an abelian $\com$-algebra sheaf on
it\footnote{Which makes the doublet $(X,\A)$ a $\mathbf{C}$-{\em
algebraized space} built into the differential triad.}, $\modl$ an
$A$-module sheaf on it with $E$ usually taken to be the
$\Z_{+}$-graded $A$-module
$\Omega=\bigoplus_{i}\Omega^{i}$\footnote{With the sheaf of
$\Omega$s denoted by boldface $\diff$.} of (complex) differential
forms, and $\bkd$ is a Cartan-K\"{a}hler type of differential
operator effecting $\modl$-subsheaf morphisms of the following
sort: $\bkd:~\mathbf{\Omega^{i}}\rightarrow\mathbf{\Omega^{i+1}}$
\cite{malrap}\footnote{The reader is referred to expression
(\ref{eq1}) below where such a differential triad is put into
cohomological liturgy.}.

\subsection{About the axioms}

Essentially, the ADG theory is based on the following two axioms
or assumptions:

\begin{itemize}

\item (a) The following (abstract) de Rham complex

\begin{equation}\label{eq1}
\begin{array}{c}
\mathbf{0}(\equiv\mathbf{\Omega^{-2}})\stackrel{\imath\equiv
d^{-2}}{\mapto}\mathbf{C}(\equiv\mathbf{\Omega^{-1}})\stackrel{\epsilon\equiv
d^{-1}}{\mapto}\mathbf{\A}(\equiv\mathbf{\Omega^{0}})\cr
\stackrel{d^{0}\equiv\partial}{\mapto}\mathbf{\Omega^{1}}\stackrel{d^{1}\equiv\kd}{\mapto}
\mathbf{\Omega^{2}}\stackrel{d^{2}}{\mapto}\cdots\mathbf{\Omega^{n}}
\stackrel{d^{n}}{\mapto}\cdots
\end{array}
\end{equation}

\noindent associated with the differential triad $(X,\A
,\modl\equiv\diff)$ {\em is exact}; where in (\ref{eq1}),
$\mathbf{C}$ is the constant sheaf of complex numbers $\com$, $\A$
is a commutative $\com$-algebra (structure) sheaf and the
$\mathbf{\Omega^{i}}$s are (sub)sheaves (of $\diff$) of
($\Z_{+}$-graded and complex) differential $\A$-modules. As we
also mentioned in the previous subsection, the $d^{i}$-arrows
($i\geq 1$) linking the sheaves in the cochain expression
(\ref{eq1}) are {\em sheaf morphisms} and, in particular,
$d^{1}\equiv\bkd$ is a nilpotent Cartan-K\"{a}hler-like
differential operator\footnote{Note also that our symbolism
$d^{-2}$ for the canonical injection $\imath$ of the trivial
constant zero sheaf $\mathbf{0}\equiv\mathbf{\Omega^{-2}}$ into
$\mathbf{C}$, and $d^{-1}$ for the canonical embedding $\epsilon$
of the complex numbers into the structure algebra sheaf
$\A\equiv\mathbf{\Omega^{-1}}$, is a non-standard one not to be
found in either \cite{mall1} or \cite{mall2}. It was adopted for
`symbolic completeness' and clarity. Finally,
$d^{0}\equiv\partial$ is the usual partial differential operator
acting in the usual way on the abelian `coordinate
$\com$-algebras' dwelling in the stalks of the structure sheaf
$\A$.}.

\item (b) There is a {\em short exact exponential
sheaf sequence}\footnote{In fact, it should also be mentioned here
that (a) entails any {\em short exact sequence} from (\ref{eq1})
\cite{mall1, mall2}.}.

\end{itemize}

In the present paper we are only going to deal in detail with
axiom (a) (section \ref{sec4}) since, as it was discussed in
section \ref{sec1}, we would like to study `purely cohomological'
features of the finsheaves in \cite{malrap}; hence, we leave the
relatively secondary assumption (b) (and the one mentioned in the
footnote following it) to the reader's curiosity which, however,
can be amply satisfied from reading \cite{mall1, mall2}. We must
also remark here that, since we wish to apply ADG to the finitary
regime (sections \ref{sec4}--\ref{sec7}), the assumption (a) above
is not an `axiom' proper ({\it ie}, a primitive assumption) any
more; rather, it is a proposition (about the exactness of the de
Rham complex) that we must actually show that it holds true in the
locally finite case. We argue for this in subsection 4.2.

\subsection{About the technique}

We read from \cite{mall2} that the main technique employed in the
{\it aufbau} of ADG is {\em sheaf-cohomology}\footnote{The reader
is also referred to \cite{strooker} for a nice and relatively
`down to earth' introduction to sheaf-cohomology from a modern
categorical perspective.}. It is fair to say that the first
author's main mathematical motivation for building ADG was the
possibility of abstracting, and concomitantly generalizing, the
usual de Rham cohomology of the `classical' differential calculus
on $\smooth$-smooth manifolds by using sheaf-cohomological
techniques on vector, algebra and (differential) module sheaves
over relatively arbitrary topological spaces with ultimate aim
towards resolving, or even possibly evading altogether, the
singular smooth manifold theory when viewed from a broader and
more potent sheaf-theoretic perspective\footnote{See section
\ref{sec1}.}. Such an endeavor, that is, to generalize the usual
de Rham theory, is most welcome also from a physical point of view
since, and we quote von Westenholz from \cite{west}, ``{\itshape
the structure underlying an intrinsic approach to physics is
`essentially' de Rham-cohomology}''. At the same time, we may
recall Wheeler's fundamental insight that in the higgledy-piggledy
realm of the quantum perhaps the sole operative `principle' is one
of ``{\itshape law without law}'', which, in turn, can also be
translated in (co)homological terms to the by now famous motto
``{\itshape the boundary of a boundary is zero}'' \cite{wheel},
and it is well known that the latter lies at the heart of (de
Rham) cohomology and, as we will see in the present work, in the
latter's sheaf-theoretic generalization by ADG.

Now, on to a few slightly more technical details: a central notion
in the sheaf-cohomology used in ADG is that of an {\em
$\A$-resolution of an abstract $\A$-module sheaf $\modl$}, by
which one means any cochain $\A$-complex of positive degree or
grade

\begin{equation}\label{eq2}
\mathcal{S}^{^\centerdot}:~~\mathbf{0}\mapto\mathbf{\mathcal{S}^{0}}\stackrel{d^{0}}{\mapto}
\mathbf{\mathcal{S}^{1}}
\stackrel{d^{1}}{\mapto}\mathbf{\mathcal{S}^{2}}\stackrel{d^{2}}{\mapto}\cdots
\end{equation}

\noindent securing that the following `$\mathbf{\modl}$-enriched'
$\A$-complex

\begin{equation}\label{eq3}
\tilde{\mathcal{S}}^{^\centerdot}:~~\mathbf{0}\mapto\mathbf{\modl}\stackrel{\imath}{\mapto}
\mathbf{\mathcal{S}^{0}}\stackrel{d^{0}}{\mapto}\mathbf{\mathcal{S}^{1}}
\stackrel{d^{1}}{\mapto}\mathbf{\mathcal{S}^{2}}\stackrel{d^{2}}{\mapto}\cdots
\end{equation}

\noindent {\em is exact}. More particularly, if the
$\mathbf{\mathcal{S}^{i}}$s in (\ref{eq3}) are injective
$\A$-modules, the $\A$-resolution of $\mathbf{\modl}$ is called
`{\em injective}'. In fact, any given $\A$-module sheaf $\modl$
(on an arbitrary base topological space $X$) admits an injective
resolution {\it \`a la} (\ref{eq3}) \cite{mall1}. Such injective
resolutions are of great import in defining non-trivial
sheaf-cohomological generalizations (or abstractions) of the
concrete de Rham complex on a $\smooth$-smooth
manifold\footnote{This is obtained by identifying the structure
(coordinate) algebra sheaf $\A$ in (\ref{eq1}) with the algebra
sheaf ${}^{\com}\mathbf{\smooth}(X)$ of infinitely differentiable
$\com$-valued functions on $X$ ({\it ie}, $\A(X)\equiv
{}^{\com}\mathbf{\smooth}(X)$).} and its corresponding de Rham
theorem. We return to them and their use in more detail in section
\ref{sec4}.

\subsection{About the philosophy}

We also read from \cite{mall2} that the philosophy underlying a
sheaf-theoretic approach to differential geometry, with its
intrinsic, abstract in nature, sheaf-cohomological mechanism, is
of an algebraic-operationalistic character. This seems to suit the
general philosophy of quantum mechanics according to which what is
of physical importance, the `physically real' so to speak, is less
the classical ideal of `background absolute objects' (such as
`spacetime', for instance) existing `inertly out there'
independently of ({\it ie}, not responding to) our own dynamically
perturbing operations of observing `them', and more these
operations (or dynamical actions) themselves \cite{df96}---which
operations, in turn, can be conveniently organized into algebra
sheaves \cite{rap2, malrap, rap3}. In a nutshell, ADG has made us
realize that space(time) (especially the classical, pointed
geometric $\smooth$-smooth model of it) is really of secondary
importance for doing differential geometry; while, in practice, of
primary importance are the algebraically represented (dynamical)
relations between objects living on this space---which space,
especially in its continuous guise, we actually do not have
experience of anyway \cite{rapzap1, rapzap2}. In view of this
undermining of the smooth spacetime continuum that we wish to
propound here, it is accurate to say that the central didactic
point learned from ADG is that one should in a sense turn the
tables around and instead of using algebras of $\smooth$-functions
to coordinatize (as it were, to measure!) space(time) when, as a
matter of fact, these very algebras derive from the differential
manifold space itself, one should rather commence with a structure
algebra sheaf $\A$ suitable to one's physical problem and {\em
derive space(time)} and possibly its (differential) geometric
features from it. Algebra (ultimately, dynamics) comes first;
while, space and its (differential) geometric properties second.
At the same time, this `manifold neglect' that we advocate here is
even more prominent in current quantum gravity research where the
classical ideal of an inert, fixed, absolute, ether-like
background geometrical smooth spacetime continuum \cite{einst24},
with its endogenous pathologies and unphysical infinities, should
arguably be replaced by something of a more reticular-algebraic
and dynamical character \cite{einst36, einst56, crane95}. We
strongly feel that sheaf theory, especially in the intrinsically
algebraic manner used by ADG, can provide a suitable language and
useful tools for developing such an entirely algebraic description
of quantum reality, in particular, of quantum spacetime structure
and its dynamics ({\it ie}, quantum gravity) \cite{malrap}.

On a more modest note, to these authors the present paper will
have fulfilled a significant part of its purpose if it introduced
and managed to make plain to a wider readership of (mathematical)
physicists---in particular those interested in or actually working
on quantum gravity---some central concepts, constructions and
results from ADG, as well as how they may prove to be useful to
their research. We also believe that the application of ideas from
ADG to a particular finitary-algebraic context and to its
associated discrete Lorentzian quantum gravity research program as
in the present paper, will further enhance the familiarization of
the reader with the basic notions and structures of the abstract
theory developed in \cite{mall1, mall2, mall4}. Our locally
finite, causal and quantal version of Lorentzian gravity
\cite{malrap, rap6} may be regarded as a physical toy model of
ADG.

\section{Rudiments of finitary \v{C}ech (co)homology}\label{sec3}

In this section we present the basic elements of a finitary
version of the usual \v{C}ech cohomology of a
(paracompact\footnote{One may wish to recall that a topological
space $X$ is said to be paracompact if every open cover of it
admits a locally finite refinement \cite{dugu}.}) $\smooth$-smooth
manifold. The epithet `finitary' pertains to a particular
procedure or `algorithm', due to Sorkin \cite{sork91}, for
substituting bounded regions $X$\footnote{$X$ is said to be
bounded when its closure is compact. Such a space is otherwise
known in the mathematical literature as `relatively compact'
\cite{dugu}.} of $\mathcal{C}^{0}$-manifolds
$M$\footnote{Technically speaking, $M$ is said to be
$\mathcal{C}^{0}$-continuous when it is a topological manifold.}
by partially ordered sets (posets) relative to $X$'s {\em locally
finite open covers}\footnote{An open cover or local frame $\gauge$
of $X$ is said to be locally finite if every point $x$ of $X$ has
an open neighborhood that intersects a finite number of the
covering open sets $U$ in $\gauge$. Ultimately, every point $x$ of
$X$ belongs to a finite number of open sets in the covering
$\gauge$.}. We further restrict our attention to `finitary poset
substitutes' of $X$ that are {\em simplicial complexes}
\cite{rapzap1, zap1, rapzap2}. Such posets, for instance, are the
ones obtained from the so-called `{\em nerve construction}'
originally due to \v{C}ech \cite{eilsteen} and subsequently, quite
independently, due to Alexandrov \cite{alex1, alex2}. These
finitary ({\it ie}, locally finite) simplicial skeletonizations of
$\mathcal{C}^{0}$-manifolds will provide the essential homological
backbone on which we are going to support their dual finitary
Rota-algebraic \v{C}ech cohomological elaborations in the sequel.
So, let us commence with Sorkin's algorithm.

\subsection{Finitary $\mathbf{\mathcal{C}^{0}}$-substitutes
revisited}

Below, we briefly review Sorkin's recipe for replacing
$\mathcal{C}^{0}$-spacetime continua by finitary $T_{0}$-poset
topological spaces\footnote{Recall that a topological space is
said to be $T_{0}$ if for every pair $(x,y)$ of points in it
either $x$ or $y$ possesses an open neighborhood about it that
does not include the other ({\it ie}, $\forall x,y\in X,~~ \exists
\mathcal{O}(x)~ {\rm
or}~\mathcal{O}(y):~~y\not\in\mathcal{O}(x)~{\rm
or}~x\not\in\mathcal{O}(y)$) \cite{dugu}.} relative to locally
finite open coverings. The original algorithm can be found in
$\cite{sork91}$, and in less but sufficient detail in
\cite{rapzap1, rap1, rap2, malrap, rapzap2}.

Let $X$ be a bounded region in a topological manifold $M$. Assume
that $\gauge=\{ U\}$ is a locally finite open cover or
`coordinatizing frame' (or even `local gauge basis')\footnote{See
again subsection 2.1.} of $X$. For every point $x$ of $X$
symbolize by $\Lambda(x)$ the smallest open neighborhood covering
$x$ in the subtopology $\mathcal{T}$ of $X$ generated by
$\gauge$\footnote{The subtopology $\mathcal{T}(\gauge)$ of $X$ is
generated by arbitrary unions of finite intersections of the open
$U$s in $\gauge$ ({\it viz.}, in other words, `the topology on the
set $X$ generated by $\gauge$' or `the open sets in $\gauge$
constitute a sub-basis for the topology $\mathcal{T}$ of $X$', the
latter being, by the hypothesis for $\gauge$, weaker than the
initial $\mathcal{C}^{0}$-manifold topology on $X$
\cite{sork91}.}: $\Lambda(x)|_{\gauge}:=\bigcap\{ U\in\gauge:~
x\in U\}$. Then, define the following preorder relation
`$\rightharpoonup$'\footnote{We recall that a preorder is a
reflexive and transitive binary relation.} between $X$'s points
with respect to their $\Lambda$s

\begin{equation}\label{eq4}
x\rightharpoonup y\Leftrightarrow \Lambda(x)\subset\Lambda(y)
\end{equation}

\noindent and enquire under what condition the
`preorder-topological spaces' defined by
`$\rightharpoonup$'\footnote{The relatively discrete topology
$\mathcal{T}$ that `$\rightharpoonup$' defines is based on open
sets of the form $\mathcal{O}(x)=\{ y:~ y\rightharpoonup x\}$ and
the preorder relation $x\rightharpoonup y$ between $X$'s points
can be literally interpreted as ``{\itshape the constant sequence
$(x)$ converges to $y$ in $\mathcal{T}$}'' \cite{sork91}. For
instance, continuous maps on the preorder-topological space are
exactly the ones that preserve `$\rightharpoonup$' ({\it ie},
precisely the maps that preserve the convergence of the aforesaid
sequences!).} are $T_{0}$. We read from \cite{sork91} that this is
so when `$\rightharpoonup$' is actually a partial order
`$\rightarrow$'\footnote{Recall that a partial order is a preorder
that is also antisymmetric ({\it ie}, $(x\rightarrow
y)\wedge(y\rightarrow x)\Rightarrow x=y$).}. In order to convert
the aforementioned preorder-topological space into a
$T_{0}$-poset, one simply has to factor $X$ by the following
equivalence relation defined relative to `$\rightharpoonup$' and
$\gauge$ as

\begin{equation}\label{eq5}
x\stackrel{\gauge}{\sim}y\Leftrightarrow
\Lambda(x)=\Lambda(y)\Leftrightarrow (x\rightharpoonup
y)\wedge(y\rightharpoonup x)
\end{equation}

\noindent so that the resulting space
$P(\gauge):=X/\stackrel{\gauge}{\sim}$, consisting of
$\stackrel{\gauge}{\sim}$-equivalence classes of the points of
$X$, is a poset $T_{0}$-topological space. This so-called finitary
substitute of $X$ \cite{sork91}, $P(\gauge)$, we will henceforth
refer to as `finitary topological poset' (fintoposet) \cite{rap1}.

\subsection{\v{C}ech-Alexandrov nerves: finitary simplicial complexes}

In this subsection we present the fintoposets obtained by Sorkin's
algorithm above from a homological perspective, that is, as
simplicial complexes. This presentation is based on a well known
construction due to \v{C}ech \cite{eilsteen} and Alexandrov
\cite{alex1, alex2}, usually coined `the nerve-skeletonization of
a topological manifold relative to an open cover of it'---the
particular case of interest here being the nerve of a locally
finite open covering $\gauge$ of $X$\footnote{The reader can also
refer to \cite{dugu} for a nice introduction to the homological
nerve construction.}. Thus, an appropriate denomination for the
relevant homology theory, also in keeping with the jargon of the
fintoposet-discretizations of $\mathcal{C}^{0}$-manifolds due to
Sorkin, would be `\v{C}ech-Alexandrov finitary homology'.

The specific approach to the simplicial decompositions of
topological manifolds due to \v{C}ech-Alexandrov to be presented
below is taken mainly from \cite{rapzap1, rapzap2}. In order to be
able to apply concepts of simplicial homology to posets like the
fintoposets of the previous subsection, we give a relatively
non-standard definition of simplicial complexes deriving from the
\v{C}ech-Alexandrov nerve construction alluded to above that
effectively views them as posets. Such a definition will also come
in handy in our presentation of the dual finitary-algebraic
cohomological theory in the next subsection.

Thus, we first recall that the nerve $\mathcal{N}$ of a (finitary)
open cover $\gauge$ of the bounded region $X$ of a
$\mathcal{C}^{0}$-manifold $M$ is the simplicial complex having
for vertices the elements of $\gauge$ ({\it ie}, the covering open
sets) and for simplices subsets of vertices with non-empty
intersections. In particular, by a $k$-simplex $\mathcal{K}$ in
$\mathcal{N}$ one understands the following set of non-trivially
intersecting vertices $\{U_{0}, \ldots , U_{k}\}$

\begin{equation}\label{eq6}
\mathcal{K}=\{ U_{0}, \ldots , U_{k}\}\in\mathcal{N} \:
\Leftrightarrow \: U_{0}\cap U_{1}\cap\ldots\cap
U_{k}\neq\emptyset
\end{equation}

\noindent Now, the nerve $\mathcal{N}$ of a (locally finite) open
cover $\gauge$ of $X$, being a simplicial complex, can also be
viewed as a poset $P$---much like in the sense of Sorkin discussed
in the previous subsection\footnote{Hence our use of the same
symbol $P$ for the (finto)posets involved in both poset
constructions. Indicatively, we just note in this respect that the
basic $\Lambda(x)$ involved in Sorkin's algorithm is nothing else
but the nerve simplex of $x$ in $\gauge$ ({\it ie}, the open
set---the smallest, in fact---in the subtopology
$\mathcal{T}(\gauge)$ of $X$ obtained by the intersection of all
the $U$s in $\gauge$ that cover $x$---the latter collection, in
case $k$ open subsets of $X$ in $\gauge$ contain $x$, being a
$k$-simplex in the sense described above).}. The points of $P$ are
the simplices of the complex $\mathcal{N}$, and the partial order
arrows `$\rightarrow$' are drawn according to the following
simplicial `face rule'

\begin{equation}\label{eq7}
p \rightarrow q \: \Leftrightarrow \: p \mbox{ is a face of } q
\end{equation}

\medskip

As in \cite{zap2, rapzap1, zap1, rapzap2}, we note that in the
non-degenerate cases, the posets associated with the
\v{C}ech-Alexandrov simplicial nerves and those derived from
Sorkin's algorithm are the same. We have chosen the homological
path of nerves, because their specific algebraic structure will
make it possible to build the dual algebraic theory for Sorkin's
fintoposets via their so-called incidence Rota algebras in the
next subsection. In turn, the latter will enable us to catch
glimpses of important for our study here finitary differential and
cohomological attributes that these algebras (and the finsheaves
thereof) possess \cite{rapzap1, malrap, zap1, rapzap2}.

\subsection{The `Gelfand dual' algebraic theory: \v{C}ech-type of
cohomology on finitary spacetime sheaves of incidence algebras}

Our casting Sorkin's fintoposets in homological terms, that is, as
simplicial complexes, will prove its worth in this subsection.

First, we recall from \cite{zap2, rapzap1, rap1, malrap, rapzap2}
how to pass to algebraic objects dual to those finitary simplicial
complexes. Such finite dimensional algebras are called incidence
algebras and, in the context of enumerative combinatorics
\cite{rstan}, they were first championed by Rota
\cite{rota}\footnote{For a beautiful introduction to incidence
algebras, especially those associated with locally finite posets
that are of particular interest to us here, the reader is referred
to \cite{odspieg}.}.

With every fintoposet $P$ its incidence Rota algebra $\Omega(P)$
can be associated, as follows: first represent the arrows
$p\rightarrow q$ in $P$ in the Dirac operator ({\it ie}, ket-bra)
notation as $\ketbra{p}{q}$. Then define $\Omega(P)$, as a (finite
dimensional) $\com$-linear space, by $\Omega(P):=\spn_{\com}\{
\ketbra{p}{q}:~p\rightarrow q\in P\}$, and subsequently convert it
to a non-abelian $\com$-algebra by requiring closure under the
following non-commutative poset-categorical (semigroup) arrow
product

\begin{equation}\label{eq8} \ketbra{p}{q} \cdot \ketbra{r}{s} = \ket{p}
\braket{q}{r} \bra{s} = \braket{q}{r} \cdot \ketbra{p}{s} =
\left\lbrace
\begin{array}{rcl} \ketbra{p}{s} &,& \mbox{if } q=r \cr 0 &&
\mbox{otherwise}
\end{array} \right.
\end{equation}

\noindent which closes and is associative precisely because of the
transitivity of the partial order `$\rightarrow$' in $P$.

Using the fact that Sorkin's fintoposets are simplicial complexes
naturally characterized by a positive integer-valued grade or
degree (or even homological dimension
\cite{zap1}\footnote{Actually, the homological dimension of a
simplicial complex equals to its degree minus one.}), one can
easily show that the corresponding $\Omega(P)$s are $\Z_{+}$-{\em
graded linear spaces} \cite{rapzap1, zap1}. With respect to this
grading then, $\Omega(P)$ splits into the following direct sum of
vectors subspaces

\begin{equation}\label{eq9}
 \Omega(P) = \bigoplus_{i\in\Z_{+}}\Omega^{i}=\Omega^0 \oplus \Omega^1 \oplus
 \ldots:=A\oplus\mathcal{R}
\end{equation}

\noindent with $A:=\Omega^{0}=\spn_{\com}\{\ketbra{p}{p}\}$ a
commutative subalgebra of $\Omega(P)$ consisting of its grade zero
elements\footnote{Again, we read from \cite{rapzap1, rap1, zap1}
that this abelian subalgebra is symbolized by $\aconn$, but in the
present study, as also alluded to in the previous section, we
reserve this symbol for the spin-Lorentzian connections, and
rather use $A$ for such algebras (and $\A$ for sheaves of them).},
and $\mathcal{R}:=\bigoplus_{i\geq1}\Omega^{i}$ a linear subspace
of $\Omega(P)$ spanned over the $\com$ by elements of grade
greater than or equal to one.

The crucial fact is that the correspondence
$P\rightarrow\Omega(P)$ is the object-wise part of a {\em
contravariant functor} from the category $\mathfrak{P}$ of
fintoposets and order morphisms ({\it ie}, `fincontinuous' or
`$\rightarrow$'-monotone maps \cite{sork91})\footnote{Or its
equivalent category consisting of simplicial complexes and
simplicial maps. In \cite{rap3} $\mathfrak{P}$ was coined `the
Alexandrov-Sorkin poset category'. Here we may add \v{C}ech's
contribution to it and call it `the \v{C}ech-Alexandrov-Sorkin
category'.} to the category $\mathfrak{R}$ of incidence algebras
and algebra homomorphisms \cite{zap1}\footnote{In \cite{rap3}
$\mathfrak{R}$ was called `the Rota-Zapatrin category'.}, thus it
is a categorical sort of duality. In fact, the very `{\em Gelfand
spatialization}' procedure employed in \cite{zap1, brezap,
rapzap1, rap1} in order to assign a topology onto the (primitive
spectra consisting of kernels of equivalence classes of
irreducible representations of the) $\Omega(P)$s in such a way
that they are {\em locally homeomorphic} to the fintoposets $P$
from which they were derived\footnote{The reader should keep this
remark in mind for what follows.} \cite{malrap, rap3}, was
essentially based on this categorical duality between fintoposets
and their incidence algebras. From now on we will refer to it as
`{\em Gelfand duality}'.

It is also precisely due to the Gelfand duality between
$\mathfrak{P}$ and $\mathfrak{R}$ that Zapatrin was able to first
develop a sound {\em homological} theory for fintoposets or their
equivalent \v{C}ech-Alexandrov nerves in $\mathfrak{P}$, and then
to translate it to a {\em cohomological} theory for their
corresponding incidence algebras in $\mathfrak{R}$ \cite{zap1}.
For instance, a Cartan-K\"{a}hler-type of nilpotent differential
operator $\bkd$---arguably {\em the} operator to initiate a
cohomological treatment of the $\Omega(P)$s in
$\mathfrak{R}$---was constructed (implicitly by using the Gelfand
duality) from a suitable finitary version of the homological
border (boundary) $\delta$ and coborder (coboundary) $\delta^{*}$
operators acting on the objects of $\mathfrak{P}$.

Indeed, with the definition of $\bkd$ one can straightforwardly
see that the $\Omega(P)$s in (\ref{eq9}) are $A$-{\em modules of}
$\Z_{+}$-{\em graded discrete differential forms} \cite{rapzap1,
zap1, rapzap2}, otherwise known as {\em discrete differential
manifolds} \cite{dim1, dim2}. In particular, $\Omega(P)$'s abelian
subalgebra consisting of scalar-like quantities,
$A\equiv\Omega^{0}$, corresponds to a reticular version of the
algebra ${}^{\com}\smooth(X)$ of $\com$-valued smooth coordinates
of the classical manifold's point events, while its linear
subspace $\mathcal{R}$ over $A$ to a discrete version of the
graded $A$-bimodule of differential forms cotangent to every point
of the classical (complex) $\smooth$-smooth manifold
\cite{rapzap1}\footnote{See also \cite{brezap1}.}. The action of
$\bkd$ is to effect transitions between the linear subspaces
$\Omega^{i}$ of $\Omega(P)$ in (\ref{eq9}), as follows: $\bkd
:~\Omega^{i}\rightarrow\Omega^{i+1}$ \cite{dim2, dim1, rapzap1,
brezap1, zap1, malrap}. All in all, the bonus from studying the
finite dimensional incidence algebraic (cohomological) objects in
$\mathfrak{R}$ which are Gelfand dual to the fintoposet/simplicial
complex (homological) objects in $\mathfrak{P}$ is that the former
encode information, in an inherently discrete guise, not only
about the continuous-topological ({\it ie}, the $\mathcal{C}^{0}$)
structure of the classical spacetime manifold like their dual
correspondents $P$ in $\mathfrak{P}$ do \cite{sork91}, but also
about its differential ({\it ie}, the $\smooth$) structure
\cite{rapzap1, malrap, rapzap2}.

Furthermore, now that we have a sort of exterior derivative
operator $\bkd$ in our hands, all that we need to actually
commence a finitary \v{C}ech-type of sheaf-cohomological study of
our reticular-algebraic structures is to organize the incidence
algebras into algebra sheaves and then apply to the latter ideas,
techniques and results from Mallios' ADG \cite{mall1, mall2}. To
this end, we recall first briefly the notion of {\em finitary
spacetime sheaves} (finsheaves) from \cite{rap2} and then the
finsheaves of incidence algebras from \cite{malrap}.

In \cite{rap2}, finsheaves of $\mathcal{C}^{0}$-observables of the
continuous topology of a bounded region $X$ of a topological
spacetime manifold were defined as function spaces that are {\em
locally homeomorphic}\footnote{See discussion around footnote 40
above.} to the base fintoposet substitutes of the locally
Euclidean manifold topology of $X$ thus, technically speaking,
{\em sheaves} over them \cite{mall1, mall2}. Subsequently in
\cite{malrap}, the stalks\footnote{The stalk of a sheaf is
more-or-less analogous to the fiber space of a fiber bundle---it
is the point-wise (relative to the topological base space on which
the sheaf is soldered) local structure of the sheaf space. For
instance, as a non-topologized set, a sheaf $\mathcal{S}$ over a
topological space $X$, $\mathcal{S}(X)$, carries the discrete
topology of its stalks point-wise over $X$, as:
$\mathcal{S}(X)=\bigoplus_{x\in X}\mathcal{S}_{i}$; where
$\mathcal{S}_{i}$ are its stalks and the direct sum sign may also
be thought of as the disjoint union operation.} of those
$\mathcal{C}^{0}$-finsheaves were endowed with further algebraic
structure in a way that this extra structure respects the
horizontal (local) `fincontinuity' ({\it ie}, the finitary
topology) of the base fintoposets---thus, ultimately, it respects
the sheaf structure itself \cite{mall1, mall2}.

More specifically, finsheaves of incidence Rota algebras over
Sorkin's fintoposets were defined in \cite{malrap}. We may
symbolize these by $\diff(P)$ and, as said before, omit the
finitary base topological space $P$ from its argument unless we
would like to comment on it. Incidence algebras $\Omega$ dwell in
the stalks of $\diff$ and the (germs of continuous) sections of
the latter\footnote{The germs of continuous sections of a sheaf
$\mathcal{S}$ by definition take values in its stalks.} inherit
the algebraic structure of the $\Omega$s for, after all, ``{\em a
sheaf} (of whatever algebraic objects) {\em is its sections}''
\cite{mall1, mall2}\footnote{This gives a pivotal role to the
notion of `section of a sheaf' in Mallios' ADG, as we will also
witness in the sequel.}. Furthermore, $\bkd$ lifts in $\diff$ to
effect transitions between its $\Z_{+}$-graded
$\mathbf{\Omega^{i}}$ vector subsheaves, in the following manner:
$\bkd:~\mathbf{\Omega^{i}}\rightarrow\mathbf{\Omega^{i+1}}$.

For the finsheaf-cohomological aspirations of the present study we
note that the triplet
$\mathcal{T}_{m}:=(P_{m},\mathbf{\Omega_{m}},\bkd)$\footnote{The
subscript `$m$' is the so-called `finitarity or resolution index'
and its (physical) meaning can be obtained directly from
\cite{sork91, rap2, malrap}. We will use it in section
\ref{sec5}.} is a finitary version of the classical ({\it ie},
$\smooth$-smooth) differential triad
$\mathcal{T}_{\infty}:=(X,\mathbf{\modl}\equiv
\mathbf{^{\com}\Omega}_{C},d)$\footnote{As we also said in the
previous footnote, that $(X,\mathbf{^{\com}\Omega}_{C},d)$ has an
infinite resolution index $n$ will be explained in \ref{sec5}.};
where $X$ is a (bounded region of a) paracompact Hausdorff
$\smooth$-smooth manifold $M$, $\mathbf{^{\com}\Omega}_{C}$ is the
sheaf of $\Z_{+}$-graded modules of Cartan's\footnote{Hence the
subscript `$C$' to the sheaf $\diff$.} (complex\footnote{This more
or less implies that one should use a complexified manifold $M$,
$^{\com}M$, and its (co)tangent bundle $T^{(*)}_{\mathbf{C}}M$
(\cite{manin}; see also subsection 6.1), but as it was also
mentioned in \cite{malrap}, here we are not going to deal with the
`$\R$ versus $\com$ spacetime debate'.}) smooth (exterior)
differential forms, and $d$ is the usual nilpotent
Cartan-K\"{a}hler (exterior) differential operator effecting
(sub)sheaf morphisms of the form:
$d:~\mathbf{\Omega^{i}}\rightarrow\mathbf{\Omega^{i+1}}$
\cite{malrap}\footnote{Interestingly enough, and in a
non-sheaf-theoretic context, Zapatrin \cite{zap1} has coined the
general triple $\mathcal{D}=(\Omega, \aconn, \kd)$---where
$\Omega$ is a graded algebra, $\aconn\equiv\Omega^{0}$ an abelian
subalgebra of $\Omega$, and $\kd$ a K\"{a}hler-type of
differential---`a differential module $\mathcal{D}$ over the basic
algebra $\aconn$'. The correspondence with our
(fin)sheaf-theoretic differential triads above is immediate: the
latter are simply (fin)sheaves of $\mathcal{D}$ in the sense of
Zapatrin. Moreover, since $\kd$ is nilpotent and we can identify
in the manner of Raptis-Zapatrin \cite{rapzap1, rapzap2} and
Zapatrin \cite{zap1}: $\kd^{0}\equiv\partial
:\Omega^{0}\mapto\Omega^{1}$ (see \ref{eq1}), as well as:
$\kd^{1}\equiv d^{1} :\Omega^{1}\mapto\Omega^{2}$ and
$\kd^{2}\equiv d^{2} :\Omega^{2}\mapto\Omega^{3}$, then the
following relations are also satisfied in the finitary regime:
$d^{1}\circ d^{0}=0=d^{2}\circ d^{1}$---a crucial condition for
the exactness of the de Rham complex in (\ref{eq1}) \cite{mall1,
mall2}.}.

In connection with the penultimate footnote however, we note that
built into the classical differential triad $\mathcal{T}_{\infty}$
is the classical $\smooth$-smooth $\mathbf{C}$-algebraized space
$(X,\mathbf{\A}\equiv\mathbf{^{\com}\smooth}(X))$ over whose
$\A$-structure sheaf's objects $A$ ({\it ie}, the algebras of
$\com$-valued $\smooth$-smooth functions on $X$) the Cartan forms
in the differential modules $\Omega^{i}$ superpose\footnote{The
reader should refresh her memory about all these technical terms
borrowed from ADG \cite{mall1, mall2} by referring back to
subsection 3.1.}. In fact, we emphasize from \cite{mall1, mall2}
that the entire differential calculus on smooth manifolds ({\it
ie}, the so-called `classical differential geometry') is based on
the assumption of $\mathbf{\A}\equiv\mathbf{^{\com}\smooth}(X)$
for structure sheaf of coordinates or
$c$-coefficients\footnote{See subsection 2.1.} of the relevant
differential triad, so that ADG's power of abstracting and
generalizing the classical calculus on smooth manifolds basically
lies in the possibility of assuming other more general or `exotic'
(in fact, possibly more singular!) coordinates ({\it ie}, local
sections of more general abelian $c$-coefficient structure sheaves
$\A$) while at the same time retaining {\em almost all} of the
innate (algebraic) mechanism and techniques of classical
differential geometry on smooth manifolds. All this was
anticipated in section \ref{sec1}.

Before we engage into some `hard core' \v{C}ech-de Rham-type of
finsheaf-cohomology on the objects inhabiting the stalks of the
vector, algebra and differential module sheaves in the finitary
differential triad $\mathcal{T}_{m}$ in the next section, we make
brief comments on the base topological spaces $P_{m}$ involved in
the $\mathcal{T}_{m}$s. These are Sorkin's fintoposets and they
are perfectly legitimate and admissible topological spaces on
which to localize the vector, algebra and module sheaves of our
particular interest and, more importantly, to perform differential
geometry {\it \`a la} ADG. For as we emphasized in section
\ref{sec2}, ADG is of such generality, and its concepts,
constructions and results of such a wide scope and applicability,
that {\em in principle it admits any topological space for base
space on which to solder the relevant sheaves and carry out
differential geometry on them} \cite{mall1, mall2}. For example,
we recall the second author's early anticipation at the end of
\cite{rap2} (where finsheaves had just been defined!) that if one
relaxed the two basic assumptions of {\em paracompactness} and
{\em Hausdorffness} (or $T_{2}$-{\em ness}\footnote{The reader may
now wish to recall that a topological space $X$ is said to be
Hausdorff, or satisfying the $T_{2}$ axiom of separation of point
set topology, if for every pair of distinct points $x$ and $y$ in
it, there exist disjoint open neighborhoods $\mathcal{O}(x)$ and
$\mathcal{O}(y)$ about them ({\it ie}, $\mathcal{O}(x)\cap
\mathcal{O}(y)=\emptyset$) \cite{dugu}.}) of ADG about the
topological character of the base spaces admissible by the theory
\footnote{See section \ref{sec2}.} to {\em relative compactness}
and $T_{1}$-{\em ness}\footnote{The reader may now like to recall
that a topological space $X$ is said to be $T_{1}$, or satisfying
the first axiom of separation of point set topology, if for every
pair $(x,y)$ of points in it {\em both} possess open neighborhoods
about them that do not include each other points ({\it ie},
$\forall x,y\in X,~~ \exists \mathcal{O}(x)~ {\rm
and}~\mathcal{O}(y):~~y\not\in\mathcal{O}(x)~{\rm
and}~x\not\in\mathcal{O}(y)$) \cite{dugu}.}---which are precisely
the two essential conditions on the $\mathcal{C}^{0}$-manifold $X$
from which fintoposets $P_{m}$ were derived by Sorkin's algorithm
in \cite{sork91}, then ideas of ADG could still apply to
finsheaves (of whatever algebraic structures) over them. This is
indeed so and, as the reader must have already noticed, it is
significantly exploited in the present work.

\section{Finitary \v{C}ech-de Rham sheaf-cohomology}\label{sec4}

This section is the nucleus of the present paper. Based on an
abstract version of the classical de Rham theorem on
$\smooth$-smooth manifolds, we entertain the possibility of a
non-trivial de Rham complex on our finitary differential triad
$\mathcal{T}_{m}=(P_{m},\diff_{m},\bkd)$. Thus, we particularize
the abstract case \cite{mall1, mall2} to our finitary regime.

\subsection{The abstract de Rham complex and its theorem}

In connection with the (injective) $\A$-resolution of an abstract
(differential) $A$-module sheaf expressions (\ref{eq2}) and
(\ref{eq3}) of section \ref{sec2}, we recall from \cite{mall1,
mall2} that the $n$-th cohomology group of an $A$-module sheaf
$\mathbf{\modl}(X)$, $H^{n}(X,\mathbf{\modl})$, can be defined via
its global sections
$\Gamma_{X}(\mathbf{\modl})\equiv\Gamma(X,\mathbf{\modl})$ as
follows

\begin{equation}\label{eq10}
H^{n}(X,\mathbf{\modl}):=R^{n}\Gamma(X,\mathbf{\modl}):=h^{n}[\Gamma(X,\mathbf{\mathcal{S}}^{^\centerdot})]:=
{\rm ker}\Gamma_{X}(d^{n})/{\rm im}\Gamma_{X}(d^{n-1})
\end{equation}

\noindent where $R^{n}\Gamma$ is the $n$-th right derived functor
of the global section functor
$\Gamma_{X}(.)\equiv\Gamma(X,.)$\footnote{It is rather obvious
that throughout the present paper we are working in the category
$\mathbf{Sh}(X)$ of sheaves (of arbitrary algebraic
structures---in particular, complex differential $A$-modules) over
$X$, and the functor $\Gamma_{X}$ acts on its sheaves and the
sheaf morphisms between them (in particular, on the differential
sheaf morphisms $d^{i}$; see (\ref{eq11}) below).}.

Correspondingly, the abstract $\A$-complex
$\mathcal{S}^{^\centerdot}$ defined by the resolution in
(\ref{eq2}) can be directly translated by the functor $\Gamma_{X}$
to the `global section $\A$-complex'
$\Gamma_{X}(\mathbf{\mathcal{S}}^{^\centerdot})$

\begin{equation}\label{eq11}
\begin{array}{c}
\Gamma_{X}(\mathcal{S}^{^\centerdot}):~~\Gamma_{X}(\mathbf{0})\mapto\Gamma_{X}(\mathbf{\mathcal{S}^{0}})
\stackrel{\Gamma_{X}(d^{0})}{\mapto}\Gamma_{X}(\mathbf{\mathcal{S}^{1}})
\stackrel{\Gamma_{X}(d^{1})}{\mapto}\cdots\cr
\cdots\stackrel{\Gamma_{X}(d^{n-1})}{\mapto}\Gamma_{X}(\mathbf{\mathcal{S}^{n-1}})
\stackrel{\Gamma_{X}(d^{n})}{\mapto}\Gamma_{X}(\mathbf{\mathcal{S}^{n}})\mapto\cdots
\end{array}
\end{equation}

\noindent which depicts the departure of the $\A$-differential
module sheaves in it from being exact ({\it ie}, the
non-triviality of the $\A(X)$-complex
$\Gamma_{X}(\mathbf{\mathcal{S}}^{^\centerdot})$) \cite{mall1,
mall2}\footnote{The reader should note here that the abstract
sheaf-cohomology advocated in ADG is principally concerned, via
$\Gamma_{X}$, with the sections of the sheaves involved, thus
vindicating and further exploiting the popular {\it motto} stated
in subsection 3.3 that `a sheaf is its sections' (see discussion
around footnote 45). Thus, in connection with the philosophy of
ADG (subsection 2.3), {\em what is of importance for ADG is more
the algebraic structure of the `objects' living on
`space(time)'---which algebraic structure, in turn, is
conveniently captured by the corresponding algebraic relations
between the sections of the respective sheaves---rather than the
underlying geometric base space(time) itself}. We would like to
thank the two Russian editors of \cite{mall1}, professors V. A.
Lyubetsky and A. V. Zarelua, for making clear and explicit in
their preface to the Russian edition of the book (vol. 1, 2000;
see footnote after \cite{mall1}) how ADG deals directly with the
geometrical objects that live on `space', thus undermining the
(physical) significance of that geometric background
`space(time)', and also how this may be of importance to current
research in theoretical physics.}. We coin
$\Gamma_{X}(\mathbf{\mathcal{S}}^{^\centerdot})$ `{\em the
abstract de Rham complex}' (ADC). We emphasize again that
$\Gamma_{X}(\mathbf{\mathcal{S}}^{^\centerdot})$ is nothing more
than the `section-wise analogue' of the abstract cochain complex
of $\com$-vector space sheaves and $\com$-linear morphisms $d^{i}$
between them that we encountered first in expression (\ref{eq1})
and subsequently in (\ref{eq2}) and (\ref{eq3})\footnote{In
\cite{mall1, mall2} for instance, the ADC in (\ref{eq1}) was
coined `the abstract de Rham complex of $X$ relative to the
differential triad
$(X,\mathbf{\Omega^{i}}(\A),d\equiv\partial)$'.}.

The ADC is the main ingredient in the expression of the abstract
de Rham theorem (ADT) which states, in a nutshell, that ``{\em the
(sheaf) cohomology of a topological space $X$, with `coefficients'
in some sheaf} $\mathbf{\modl}$ (of $\A$-modules or, more
generally, of abelian groups), {\em is that one of a certain
particular $\A$-complex} (canonically) {\em associated with the
given sheaf} $\mathbf{\modl}$; more precisely, the said cohomology
is, in fact, {\em the cohomology of any $\Gamma_{X}$-acyclic
resolution of $\mathbf{\modl}$}\footnote{The epithet `acyclic'
pertaining, of course, to the non-exactness of the ADC and the
associated non-triviality of its respective cohomology groups
$H^{n}(X,\mathbf{\mathcal{S}^{n}})$, as described above. The
abstract nature of the ADT consists in that, effectively, the
functor $\Gamma_{X}$ can be substituted by {\em any} covariant
(left exact) $\A(X)$-linear functor on $\mathbf{Sh}(X)$.}''
\cite{mall2, mall1}.

The abstract character of both the ADC and of the ADT that it
supports consists in there being generalizations of the usual
$\smooth$-smooth de Rham complex and its theorem. The reader may
like to recall that the classical de Rham theorem
(CDT)\footnote{The `C' in front of CDT could also stand for
`(c)oncrete', as opposed to the `A' (for (a)bstract) in front of
ADC and its ADT.}, which refers to the \v{C}ech cohomology of a
paracompact Hausdorff $\smooth$-smooth manifold $X$, pertains to
the cohomology of a $\Gamma_{X}$-acyclic resolution of the
constant sheaf $\mathbf{C}$ provided by the standard de Rham
complex which we bring forth from (\ref{eq1}) in a slightly
different form

\begin{equation}\label{eq12}
\Omega^{^\centerdot\infty}_{\rm de R}:~
\mathbf{0}\mapto\mathbf{\smooth}(X)\equiv\mathbf{\Omega^{0}}\stackrel{d}{\mapto}
\mathbf{\Omega^{1}}\stackrel{d}{\mapto}\mathbf{\Omega^{2}}\stackrel{d}{\mapto}
\cdots\mapto\mathbf{\Omega^{n}}\mapto 0
\end{equation}

\noindent which complex, when
$\mathbf{C}$-enriched\footnote{Recall that $\mathbf{C}$ is the
constant sheaf of the complex numbers $\com$ on $X$. Also, the
superscript `$\infty$' to $\Omega^{^\centerdot}_{\rm de R}$
reflects that we are dealing with the classical $\smooth$-smooth
case ({\it ie}, the case of infinite finitarity or resolution
index \cite{rap2, malrap}).}, provides the following exact
sequence of $\com$-vector sheaves on $X$

\begin{equation}\label{eq13}
\Omega^{^\centerdot\infty}_{\rm de R}:~
\mathbf{0}\mapto\mathbf{C}\mapto\mathbf{\smooth}(X)\equiv\mathbf{\Omega^{0}}\stackrel{d}{\mapto}
\mathbf{\Omega^{1}}\stackrel{d}{\mapto}\mathbf{\Omega^{2}}\stackrel{d}{\mapto}
\cdots\mapto\mathbf{\Omega^{n}}\mapto 0
\end{equation}

\noindent with $n$ the dimensionality of the base
$\smooth$-manifold $X$.

It is well known of course that the CDT is rooted on the lemma of
Poincar\'{e} which holds that {\em every closed $\smooth$-form on
$X$ is exact}---this statement always being true at least locally
({\it ie}, $U$-wise) in $X$. Also, we just remark here that the
acyclicity of $\Gamma_{X}$ in (\ref{eq12}) is secured by the fact
that the coordinate structure sheaf
$\A\equiv\Omega^{0}\equiv{}^{\com}\smooth(X)$ is fine on
$X$\footnote{We may recall from \cite{strooker} or \cite{mall1}
that a sheaf $\mathbf{\mathcal{S}}$ is said to be fine if for
every locally finite open covering (or every choice of
coordinatizing local gauges) $\gauge=\{U_{i}\}$ of $X$ there is a
collection of (endo)morphisms
$f_{i}:~\mathbf{\mathcal{S}}\rightarrow\mathbf{\mathcal{S}}$, such
that: (i) $\forall i, ~~|f_{i}|:=\{ x\in X:~ (f_{i})_{x}\not=
0\}\subset\overline{U_{i}}$, and (ii) $\sum_{i}f_{i}=1$ (partition
of unity).  The fineness of our finsheaves is implicitly secured
by their construction in \cite{rap2, malrap}, since, as we
mentioned earlier, the region $X$ of a $\smooth$-manifold $M$
considered there was assumed to be relatively compact ({\it ie},
bounded) \cite{sork91}, as well as that it admitted locally finite
open coverings $\gauge_{i}$; hence, {\it in extenso}, for all
practical purposes and without loss of any generality in the
construction, one could assume up-front that $X$ is, in fact, {\em
paracompact}. The latter assumption would then immediately secure
(ii) above ({\it ie}, `partition of unity') \cite{mall1}. Then,
condition (i) would also be satisfied since ``{\em every
paracompact space is normal}'' (Dieudonn\'{e}) \cite{bourb, dugu},
and `normality' for a topological space entails that {\em every
locally finite open covering of it admits a `shrinking', `precise'
refinement} \cite{dugu, mall1}. Now that we have established that
$\mathbf{\Omega^{0}_{m}}$ is fine, so are the finsheaves of graded
modules of differentials over it \cite{mall1}. The fineness of
these finsheaves will play a central role in establishing the
acyclicity of the corresponding $\Gamma_{X}$ functor on the
finitary de Rham complex in the next subsection.}.

We conclude this subsection by making the well known remark that
{\em on a paracompact $T_{2}$-space, sheaf-cohomology coincides
with the standard \v{C}ech cohomology}, and we add that it is
precisely this fact, aided by the finitary-algebraic
discretizations of $\cont$-manifolds {\it \`a la}
\v{C}ech-Alexandrov-Sorkin-Zapatrin, the finsheaves thereof
\cite{rap2, malrap} and the broad sheaf-cohomological ideas of
ADG, which technically conspired towards the conception of the
finitary \v{C}ech-de Rham cohomology presented here. We are now in
a position to present the finitary de Rham complex (FDC) and
theorem (FDT).

\subsection{The finitary de Rham complex and theorem}

We simply write the following FD
$\A\equiv\mathbf{\Omega_{m}^{0}}$-complex\footnote{We refresh the
reader's memory by noting that the subscript `$m$' here is the
finitarity or resolution index.} for the finitary differential
triad $\mathcal{T}_{m}:=(P_{m},\mathbf{\Omega_{m}},\bkd)$ defined
in the penultimate subsection

\begin{equation}\label{eq14}
\Omega^{^\centerdot m}_{\rm de R}:~
\mathbf{0}\mapto\mathbf{\Omega^{0}_{m}}\stackrel{\kd}{\mapto}
\mathbf{\Omega^{1}_{m}}\stackrel{\kd}{\mapto}\mathbf{\Omega^{2}_{m}}\stackrel{\kd}{\mapto}
\cdots\mapto\mathbf{\Omega^{n}_{m}}\mapto 0
\end{equation}

\noindent and its $\mathbf{C}$-enriched version

\begin{equation}\label{eq15}
\Omega^{^\centerdot m}_{\rm de R}:~
\mathbf{0}\mapto\mathbf{C}\mapto\mathbf{\Omega^{0}_{m}}\stackrel{\kd}{\mapto}
\mathbf{\Omega^{1}_{m}}\stackrel{\kd}{\mapto}\mathbf{\Omega^{2}_{m}}\stackrel{\kd}{\mapto}
\cdots\mapto\mathbf{\Omega^{n}_{m}}\mapto 0
\end{equation}

\noindent Both (\ref{eq14}) and (\ref{eq15}) depict the exactness
of the finitary de Rham complex $\Omega^{^\centerdot m}_{\rm de
R}$ whose $\Gamma^{m}_{P_{m}}$-acyclicity is secured by the fact
that the $\mathbf{\Omega^{0}_{m}}$-module sheaves
$\mathbf{\Omega^{i}_{m}}$ involved in it are in fact fine by
construction \cite{rap2, malrap}\footnote{See footnote 62 above.}.
This is essentially the content of the FDT\footnote{That is, that
the finitary simplicial \v{C}ech cohomology of the $P_{m}$s is
expressible in terms of the reticular differential forms
$\Omega_{m}$ living on them.}.

As it was remarked at the end of subsection 2.2, here we will
argue that the complex in (\ref{eq15}) above ({\it ie}, the
finitary version of the abstract de Rham complex in (\ref{eq1}))
is actually exact, thus, in effect, that the usual de Rham theory
of differential forms on $\smooth$-manifolds is still in force in
the locally finite regime (and not merely to be taken as an axiom
as (a) in 2.2 would {\it prima facie} seem to imply). Our argument
is an `inverse' one as we explain below:

We consider a bounded region $X$ of a $\smooth$-smooth manifold
$M$ for which the CDT is assumed to hold. Then we employ a locally
finite open gauge system $\gauge_{m}$ in the sense of Sorkin
\cite{sork91} to chart (coordinatize) $X$. Relative to
$\gauge_{m}$, as we mentioned in subsection 2.1, we extract by
Sorkin's algorithm the fintoposet $P_{m}$ \cite{sork91} and we
build the finsheaf $\mathbf{\Omega_{m}}$ \cite{rap2} of incidence
algebras $\Omega_{m}$ over it, $\mathbf{\Omega_{m}}(P_{m})$, as in
\cite{malrap}. Then, we know from \cite{sork91} that the
fintoposets form an inverse or projective system \cite{sol} poset
category (net) $\underleftarrow{\mathcal{N}}:=(P_{m},\succeq)$,
consisting of them and refinement partial order-preserving arrows
$\succeq$ between them, and which possesses an inverse or
projective limit space $P_{\infty}=\underleftarrow{\lim}P_{m}$
that is homeomorphic to $X$ as $\infty\leftarrow m$. Since the
$\Omega_{m}$s are (categorically) dual objects to the $P_{m}$s as
mentioned in subsection 2.3 \cite{zap2, rapzap1, rapzap2, zap1},
they form a direct or inductive system \cite{sol} (again a poset
category) $\underrightarrow{\mathcal{N}}:=(\Omega_{m},\preceq)$
consisting of the finite dimensional incidence Rota algebras
$\Omega_{m}$ associated with the fintoposets $P_{m}$ in
$\underleftarrow{\mathcal{N}}$ and injective algebra homomorphisms
$\preceq$ between them. Since the $\Omega_{m}$s are discrete
$\Z$-graded discrete differential manifolds, as it has been
extensively argued in \cite{rapzap1, malrap, rapzap2},
$\underrightarrow{\mathcal{N}}$ possesses an inductive limit
space, $\Omega_{\infty}=\underrightarrow{\lim}\Omega_{m}$ as
$m\rightarrow\infty$, which reminds one of the situation entailed
by a $\smooth$-smooth region $X$ in a differential manifold $M$
\cite{rapzap1, rapzap2, zap1}. The latter effectively means that
at the limit of infinite refinement of the topologies
$\mathcal{T}_{m}$ generated by (or having for bases) the
$\gauge_{m}$s, the inductive system
$\underrightarrow{\mathcal{N}}$ yields the Cartan spaces of
differential forms cotangent to every point of the
$\smooth$-smooth $X$ \cite{rapzap1, rapzap2}.

Now, our aforesaid `inverse' argument for the exactness of the
finitary de Rham complex in (\ref{eq15}) is based on the result
that {\em de Rham exactness is preserved under inductive
refinement}\footnote{Or equivalently, that {\em the inductive
limit functor is exact} \cite{sol}.} since the underlying locally
finite open covers $\gauge_{m}$ of Sorkin may be regarded as being
`good' \cite{bott}, still by providing a {\em cofinal system} in
the class of open coverings of $X$. Thus, since the exactness of
the de Rham complex is assumed to hold for the projective limit
space $X$, it also holds for the finsheaves $\mathbf{\Omega_{m}}$
of reticular differential forms soldered on Sorkin's $P_{m}$s. It
must be also mentioned here that, as one would expect,
Poincar\'{e}'s lemma holds locally for every contractible $U$ in
$\gauge_{m}$ and, in particular, for every contractible elementary
(`{\it ur}') cell $\Lambda(x)$ covering every point $x$ of $X$
(see (\ref{eq4}))\footnote{Recall that $\Lambda(x)$ is the
smallest open set in the subtopology $\mathcal{T}_{m}$ of $X$
generated by the contractible open sets $U$ in $\gauge_{m}$.},
thus it also holds locally in every $P_{m}$ of Sorkin by their
very construction \cite{sork91}.

We close this subsection by noting that adding to the
corresponding differential triads, $\mathcal{T}_{m}$ and
$\mathcal{T}_{\infty}$, their respective de Rham complexes,
$\Omega^{^\centerdot m}_{\rm de R}$ and
$\Omega^{^\centerdot\infty}_{\rm de R}$, one obtains the following
{\em finsheaf-cohomology differential tetrads}\footnote{From now
on this will be referred to as `fintetrad'.}:

\begin{itemize}
\item \underline{\bf Finitary:}
$\mathbf{\mathfrak{T}_{m}}:=(P_{m},\mathbf{\Omega_{m}},\bkd
,\Omega^{^\centerdot m}_{\rm de R})$

\item\underline{\bf $\mathbf{\smooth}$-Smooth (Classical):}
$\mathbf{\mathfrak{T}_{\infty}}:=(X,\diff,{\mathbf d}
,\Omega^{^\centerdot\infty}_{\rm de R})$
\end{itemize}

These definitions and the discussion preceding them bring us to
the following classical $\smooth$-limit construction.

\section{Classical $\smooth$-limit construction: recovering the
$\mathbf{\smooth}$-smooth \v{C}ech-de Rham complex}\label{sec5}

The contents of the present section are effectively encoded in the
following `commutative categorical limit diagram'

\[
\begin{CD}
P_{l}@>\pi_{l}^{-1}\equiv s_{l}>(1)>\mathcal{S}_{l}\equiv\mathbf{\Omega_{l}}@>\mathbf{\kd_{l}}>(2)>\Omega^{^\centerdot l}_{\rm de R}\\
@Vf_{lm}V(1^{'})V        @V\hat{f}_{lm}V(2^{'})V @V\tilde{f}_{lm}V(3^{'})V\\
P_{m}@>\pi_{m}^{-1}\equiv s_{m}>(3)>\mathcal{S}_{m}\equiv\mathbf{\Omega_{m}}@>\mathbf{\kd_{m}}>(4)>\Omega^{^\centerdot m}_{\rm de R}\\
\vdots & &\vdots & &\vdots\\
@Vf_{m\infty}V(4^{'})V        @V\hat{f}_{m\infty}V(5^{'})V @V\tilde{f}_{m\infty}V(6^{'})V\\
\lim_{m\rightarrow\infty}P_{m}\equiv
P_{\infty}\simeq\cont(X)@>\pi^{-1}\equiv
s>(5)>\lim_{m\rightarrow\infty}\mathbf{\Omega_{m}}\equiv\diff\simeq{}^{n}\!\!\bigwedge[\smooth(X)]@>\mathbf{d}>(6)>\Omega^{^\centerdot\infty}_{\rm
de R}
\end{CD}
\]

\noindent which we explain arrow-wise below:

\begin{itemize}

\item (a) \underline{\bf Arrows $\mathbf{(1)}$ and $\mathbf{(3)}$:}
The two horizontal arrows $(1)$ and $(3)$ depict the {\em local
homeomorphism} finsheaf maps $s_{l}$ and $s_{m}$, inverse to their
corresponding projection maps $\pi_{l}$ and $\pi_{m}$, from the
fintoposet base topological spaces $P_{l}$ and $P_{m}$ with
finitarity indices `$l$' and `$m$', to their respective finsheaf
spaces $\mathcal{S}_{l}$ and $\mathcal{S}_{m}$ \cite{rap2}. In
turn, as it was shown in \cite{malrap}, the latter can be
identified with the finsheaves $\mathbf{\Omega_{l}}$ and
$\mathbf{\Omega_{m}}$ of the incidence Rota algebras
$\Omega_{l}(P_{l})$ and $\Omega_{m}(P_{m})$.

\item (b) \underline{\bf Arrows $\mathbf{(2)}$ and
$\mathbf{(4)}$:} The two horizontal arrows $(2)$ and $(4)$
represent the Cartan-K\"ahler-like differential operators
$\mathbf{\bkd_{l}}$ and $\mathbf{\bkd_{m}}$ which, as said
earlier, effect graded subfinsheaf morphisms,
$\mathbf{\bkd_{l}}:~\mathbf{\Omega_{l}^{i}}\rightarrow\mathbf{\Omega_{l}^{i+1}}$
and
$\mathbf{\bkd_{m}}:~\mathbf{\Omega_{m}^{i}}\rightarrow\mathbf{\Omega_{m}^{i+1}}$,
within their respective finitary de Rham finsheaf-cohomological
complexes $\Omega^{^\centerdot l}_{de R}$ and $\Omega^{^\centerdot
m}_{de R}$.

\item (c) \underline{\bf Arrows $\mathbf{(1^{'})}$ and
$\mathbf{(2^{'})}$:} The two vertical arrows $(1^{'})$ and
$(2^{'})$ represent continuous injections, interpreted as
topological refinements, between the fintoposets $P_{l}$ and
$P_{m}$ ($P_{l}\preceq P_{m}\Leftrightarrow
f_{lm}:~P_{l}\rightarrow P_{m}$) \cite{sork91} and their
corresponding finsheaves $\mathcal{S}_{l}$ and $\mathcal{S}_{m}$
($\mathcal{S}_{l}\hat{\preceq}\mathcal{S}_{m}\Leftrightarrow
\hat{f}_{lm}:~\mathcal{S}_{l}\rightarrow\mathcal{S}_{m}$)
\cite{rap2}. That a continuous injection $f_{lm}$ ($P_{l}\preceq
P_{m}$) lifts to a similar continuous into map $\hat{f}_{lm}$
($\mathcal{S}_{l}\hat{\preceq}\mathcal{S}_{m})$ between the
finsheaf spaces over the base fintoposets $P_{l}$ and $P_{m}$ is
nicely encoded in the commutative diagram defined by the arrows
[$(1)$--$(2^{'})$, $(1^{'})$--$(3)$] above.

\item (d) \underline{\bf Arrow $\mathbf{(3^{'})}$:} The arrow $\tilde{f}_{lm}$
represents a functor carrying (sub)sheaves and their
$\mathbf{\bkd_{l}}$-morphisms in the fintetrad $\mathfrak{T}_{l}$
to their counterparts in the fintetrad $\mathfrak{T}_{m}$. In
complete analogy with $f_{lm}$ and $\hat{f}_{lm}$ above, we may
represent the corresponding functorial refinement relation
$\tilde{f}_{lm}$ between $\mathfrak{T}_{l}$ and $\mathfrak{T}_{m}$
as $\mathfrak{T}_{l}\tilde{\preceq}\mathfrak{T}_{m}$.

\item (e) \underline{\bf Arrows $\mathbf{(4^{'})}$ and
$\mathbf{(5^{'})}$:} These two arrows $f_{m\infty}$ and
$\hat{f}_{m\infty}$, as the diagram symbolically depicts, are the
maximal refinements obtained by subjecting the inverse or
projective systems (or nets)
$\mathcal{N}:=(P_{m},f_{lm}\equiv\preceq)$ and
$\hat{\mathcal{N}}:=(\mathcal{S}_{m}\equiv\mathbf{\Omega_{m}},\hat{f}_{lm}
\equiv\hat{\preceq})$ to the inverse limit (categorical limit) of
maximum (infinite) refinement (or localization! \cite{rap2,
malrap}) of the base fintoposets \cite{sork91} and their
corresponding finsheaves \cite{rap2} of incidence algebras
\cite{malrap} ({\it ie}, formally, as the refinement or resolution
index goes to infinity `$m\rightarrow\infty$', yielding:
$\lim_{m\rightarrow\infty}P_{m}\equiv P_{\infty}\simeq \cont(X)$
\cite{sork91, rapzap1, rap2} and
$\lim_{m\rightarrow\infty}\mathbf{\Omega_{m}}\equiv{}^{\com}\mathbf{\Omega_{\infty}}\equiv
{}^{n}\!\!\bigwedge[{}^{\com}\smooth(X)]$) \cite{rapzap1, malrap,
rapzap2}\footnote{The reader should note above that $\mathcal{N}$
and $\hat{\mathcal{N}}$ are the projective and inductive systems
$\underleftarrow{\mathcal{N}}$ and $\underrightarrow{\mathcal{N}}$
mentioned in the previous section, respectively. Only for
notational convenience we used the same limit symbols
`$\lim_{m\rightarrow\infty}$' (and the same refinement relations
$\preceq$) for both the inverse and the direct limit convergence
processes in $\mathcal{N}\equiv\underleftarrow{\mathcal{N}}$ and
$\hat{\mathcal{N}}\equiv\underrightarrow{\mathcal{N}}$,
respectively.}.

\item (f) \underline{\bf Arrow $\mathbf{(6^{'})}$:} The arrow
$(6^{'})$ expresses the `convergence', at the limit of infinite
resolution, of a net $\tilde{\mathcal{N}}:=(\Omega^{^\centerdot
m}_{\rm de R},\tilde{\preceq})$ of finitary de Rham sheaf-cochain
complexes and their injective functors $\tilde{f}_{lm}$ to the
classical $\smooth$-smooth de Rham complex
$\Omega^{^\centerdot\infty}_{\rm de R}$\footnote{Equivalently, and
in view of (d) above, one may think of the projective system
$\tilde{\mathcal{N}}$ as consisting of fintetrads
$\mathfrak{T}_{m}$ $\tilde{\preceq}$-nested by the functorial
injections $\tilde{f}_{lm}$ and converging at infinite refinement
to the $\smooth$-smooth sheaf-cohomological differential tetrad
$\mathfrak{T}_{\infty}$.}. This inverse limit convergence is in
complete analogy to the projective limit convergences recalled in
(f) above from \cite{sork91, rapzap1, rap2, malrap, rapzap2}.

\item (g) \underline{\bf Arrows $\mathbf{(5)}$ and
$\mathbf{(6)}$:} Arrow $(5)$ can be thought of as some kind of
`injection' or `embedding' of a $\cont$-manifold into a
$\smooth$-one ({\it ie}, the well known fact in the usual Calculus
that {\em differentiability implies continuity}, or equivalently,
that every $\smooth$-differential manifold is {\it a fortiori} a
$\cont$-topological one\footnote{In our finitary context, all this
is just to say that the incidence algebras associated with
Sorkin's fintoposets, as well as their finsheaves, encode discrete
information not only about the topological structure of
`spacetime', but also about its differentiable properties
\cite{zap2, rapzap1, brezap1, malrap, rapzap2, rap3, zap1}.}),
while the arrow $(6)$ depicts the inclusion of the sheaves
$\mathbf{\Omega^{i}}$ of smooth complex $\Z_{+}$-graded
differential forms into the smooth de Rham complex $\Omega_{de
R}^{^\centerdot\infty}$ in its corresponding classical
sheaf-cohomological differential tetrad
$\mathbf{\mathfrak{T}_{\infty}}$.

\end{itemize}

After having recovered the usual classical differential geometric
$\smooth$-smooth structures from our reticular-algebraic
substrata, we intend to initiate, at least, a sheaf-cohomological
classification {\it \`a la} ADG of the non-trivial ({\it ie},
non-flat) finitary spin-Lorentzian connections $\aconn_{m}$ that
were introduced and studied in \cite{malrap}. Such a possibility
for classifying the spin-Lorentzian connection fields $\aconn_{m}$
would amount to an effective transcription to an inherently
finitary and quantal gravitational model \cite{malrap} of the
analogous means ({\it ie}, techniques) and results ({\it ie},
theorems) for classifying smooth ({\it ie}, classical) Maxwell
fields \cite{sel83, mall1, mall2} .

\section{The abstract Weil integrality and Chern-Weil theorems: towards a
sheaf-cohomological classification of finitary spin-Lorentzian
connections $\mathbf{\aconn_{m}}$}\label{sec6}

As noted above, in the present section we will attempt to emulate
in a finsheaf-cohomological setting what is done in the classical
$\smooth$-smooth theory and entertain the possibility of assigning
a cohomology class to any reticular `closed $n$-form'---in
particular, to the (curvatures of the) finitary spin-Lorentzian
$1$-forms $\aconn_{m}$---dwelling in the relevant finsheaves in
their respective fintetrads $\mathfrak{T}_{m}$. Of great import in
such an endeavor is on the one hand ADG's achievement of
formulating abstract versions of both Weil's integrality theorem
(WIT) and of the Chern-Weil theorem (CWT) of the usual
differential geometry on smooth manifolds \cite{mall1, mall2}, and
on the other their possible transcription to the more concrete
finitary-algebraic regime of particular interest here, for it is
well known that both theorems lie at the heart of the theory of
characteristic classes of classical $\smooth$-smooth vector
bundles and sheaves. We only translate them to our
finitary-algebraic setting and, we emphasize once more, it is
precisely the abstract and quite universal character of ADG that
allows us to do this. However, before we present the aforesaid two
theorems and their finitary versions, let us briefly inform the
reader about how ADG defines and deals with `generalized
differentials', that is to say, {\em connections}, as well as how
the latter were applied to the reticular finsheaf models in
\cite{malrap}.

\subsection{A brief reminder of non-trivial finitary $\mathbf{\A}$-connections
$\mathbf{\aconn_{m}}$}

Let us recall from \cite{mall1, mall2} some basic sheaf-theoretic
facts about abstract and general $\A$-connections before we delve
into the particular finitary case of interest here.

Let $(\A ,\diff, \partial)$ be a differential triad consisting, as
usual, of the (commutative) $\com$-algebra structure sheaf $\A$
(of `generalized coordinates'), the sheaf $\diff$ of complex
$A$-modules $\Omega$ (of differential forms) and the
$\mathbf{C}$-derivation operator $\partial$ which is defined as a
{\em sheaf morphism}

\begin{equation}\label{eq16}
\partial:~\A\mapto\mathbf{\Omega}
\end{equation}

\noindent which is also

\begin{itemize}

\item (i) $\mathbf{C}$-linear between $\A$ and $\diff$
viewed as $\com$-vector sheaves, and

\item (ii) it satisfies Leibniz's product rule

\begin{equation}\label{eq17}
\partial(s\cdot t)=s\cdot\partial t+\partial s\cdot t
\end{equation}

\noindent which, in view of (i), implies that for every $\alpha$
in the constant sheaf $\mathbf{C}$: $\partial\alpha=0$, or
equivalently written as: $\partial|_{\mathbf{C}}=0$\footnote{Of
course, it is understood that the objects `$s$' and `$t$' involved
in (\ref{eq17}) are (global) sections of $\A$.}. We also note that
$\partial$ is no other than the arrow $d^{0}$ in (\ref{eq1}) which
extends to the higher grade $d^{i}$ ($i\geq1$) sheaf morphisms in
(\ref{eq1}) when the $\Omega$s in the sheaf $\diff$ are
$\Z_{+}$-graded differential modules (defining thus graded
subsheaves $\mathbf{\Omega^{i}}$ of $\diff$).

What it must be emphasized at this point, because it lies at the
heart of ADG's sheaf-theoretic approach to differential geometry,
is K\"{a}hler's fundamental insight that

\begin{quotation}

\noindent{\em every abelian unital ring admits a derivation map as
in} (\ref{eq16}) \cite{mall1, mall2}, hence it qualifies ADG as
{\em a purely algebraic picture of differential calculus}---one
without any essential dependence on a `background geometrical
space(time)'\footnote{For more on this see sections \ref{sec1} and
\ref{sec8}.}.

\end{quotation}

\noindent For the particular finitary application of ADG here, and
as it was also strongly stressed in \cite{malrap}, the finsheaves
of incidence algebras---which effectively are ring-like structures
\cite{odspieg}---naturally admit generalized derivations ({\it
viz.} connections; see below) in the spirit of K\"{a}hler quite
independently of the character of the geometric base space on
which these rings are localized. This algebraic conception of
derivation or connection is more in line with Leibniz's relational
intuition of this structure, rather than with Newton's more
spatial or geometrical one\footnote{We can briefly qualify this as
follows: one may recall that while Newton advocated a geometric
conception of derivative ({\it eg}, as measuring the slope of the
tangent line to the spatial curve which represents the graph of
the function on which this derivative operator acts), Leibniz
propounded a combinatory-relational (in effect, {\em algebraic})
notion of derivative---one that invokes no concept of (static)
ambient geometric space, but one that {\em derives from the ({\rm
possibly} dynamical) relations} between the objects involved in
the relational-algebraic structures in focus. He thus coined his
conception of differential calculus (which he ultimately perceived
as a `geometric calculus') `{\it ars combinatoria}'---{\em
combinatorial art}. In the same spirit, in ADG, with its finitary
applications here and in \cite{malrap}, derivations and their
abstractions-generalizations ({\it viz.} connections; see below)
derive from the algebraic structure of the objects (in fact, the
sections) living in the relevant (fin)sheaves (in the present
paper, the incidence algebras associated with the relational
fintoposets) and are not the idiosyncracies of any kind of
geometric space `out there' whatsoever (see also footnote 57).}.

\end{itemize}

Now, in ADG the abstraction and generalization of the
$\mathbf{C}$-derivation $\partial$ above to the notion of a
(non-trivial\footnote{The epithet `non-trivial' here pertains, as
we will mention shortly, to a connection whose curvature is
non-zero---commonly known as a `non-flat connection'.}) $\A$-{\em
connection} $\conn$ is accomplished in the following two steps:

\begin{itemize}

\item {\bf (a)}\underline{\bf Abstraction:} The abstraction of
$\partial$ to $\conn$ goes briefly as follows: first, one assumes
as above a differential triad $(\A ,\diff,\partial)$ and an
$\A$-module sheaf $\mathbf{\modl}$ on some topological space $X$;
then one defines an $\A$-{\em connection $\conn$ of} $\modl$, as a
map (in fact, again a {\em sheaf morphism})

\begin{equation}\label{eq18}
\conn :~\modl\mapto\modl\otimes_{\A}\diff
\cong\diff\otimes_{\A}\modl\equiv\diff(\modl)
\end{equation}

\noindent which again is:

\item (i) a $\mathbf{C}$-linear morphism between the $\mathbf{C}$-vector sheaves involved,
and

\item (ii) it satisfies the Leibniz rule which now reads

\begin{equation}\label{eq19}
\conn(\beta\cdot t)=\beta\cdot\conn(t)+t\otimes\conn(\beta)
\end{equation}

\noindent for any (sections) $\beta\in\A(U)$, $t\in\modl(U)$, with
$U\subset X$ open ({\it ie}, properly speaking,
$\beta\in\Gamma(U,\A)$ and $t\in\Gamma(U,\modl)$).

\item{\bf (b)} \underline{\bf Generalization:} As briefly alluded
to in the last footnote, the generalization of $\partial$ to
$\conn$ basically rests on the observation made in \cite{mall1,
mall2} that the former is a trivial, {\em flat
connection}\footnote{In a discrete context similar to the finitary
one of interest to us here and to the one studied in
\cite{malrap}, Dimakis and M\"{u}ller-Hoissen also observed that
the nilpotent Cartan-K\"ahler derivation $\partial\equiv d$ is a
flat kind of connection ({\it ie}, one whose curvature is zero).},
so that to generalize it means, effectively, to curve it. In
\cite{malrap}, for instance, the latter was accomplished by {\em
gauging} the relevant (fin)sheaves, which gauging, in turn, was
formally implemented by locally augmenting $\partial$ with a
non-zero gauge potential $1$-form $\aconn$\footnote{This entails
that the sheaf morphism $\conn$ in (\ref{eq18}) is, in effect, the
usual `$1$-form-valued assignment': $\conn
:~\A\mapto\mathbf{\Omega^{1}}\subset\diff$---the familiar
structure encountered in the standard vector bundle models of
gauge theories. See also below.}, as follows

\begin{equation}\label{eq20}
{\rm Formal}\,\,{\rm gauging}:~\partial\mapto\conn
=\partial+\aconn\footnote{In fact, at least locally, $\conn$ is
uniquely determined by what physicists call `the vector potential'
$\aconn$. That is why in \cite{malrap} we did not distinguish (at
least locally) between $\conn$ and its non-flat part $\aconn$.}
\end{equation}

\noindent We thus arrive at how physicists normally interpret a
connection $\conn$ as a {\em covariant derivative} which is a
result of the process of {\em gauging} or {\em localizing} a
physical structure (and its symmetries) \cite{malrap}. From the
general perspective of a non-flat connection $\conn$, the flat
case $\partial$ is a special case recovered exactly by setting
$\aconn=0$\footnote{But as it was emphasized in \cite{malrap},
from ADG's perspective \cite{mall1, mall2}, $\partial$ is a
perfectly legitimate connection; albeit, a flat or trivial one.}.
Sheaf-theoretically speaking, the process of gauging or localizing
means essentially that {\em the sheaves involved do not admit
global sections} or equivalently, and perhaps more geometrically,
the particular coordinate structure algebra sheaf $\A$ is
localized relative to the open coordinate local gauges $U$ in
$\gauge$ covering $X$ \cite{mall1, mall2}. Our generalized
coordinatizations or measurements of the loci of events in $X$, as
encoded in $\A$, are localized relative to the $U$s in $\gauge$.
In turn, on this fact we based a finitary version of the Principle
of Equivalence of general relativity on a smooth manifold and the
concomitant curving of the finsheaves of incidence algebras
modelling qausets in \cite{malrap}. We will return to this
subsequently and in the next section\footnote{It is worth
mentioning here that the $\A$-connection $\conn$ to which
$\partial$ is abstracted and generalized by (a) and (b) above, is
in complete analogy to, and we quote Kastler from \cite{kastle},
``{\em the most general notion of linear connection $\nabla$}''
used in Connes' popular Noncommutative Differential Geometry (NDG)
theory \cite{connes}. However, the epithet `noncommutative' in
Connes' work, and in contradistinction to the first author's ADG,
pertains to also admitting {\em non-abelian} structure
$\com$-algebra sheaves. In view also of the noncommutative ideas
in the context of qausets expounded in \cite{rap3}, it would
certainly be worthwhile to try to relate Connes' NDG with ADG}.

\end{itemize}

\medskip

We can now make the following three remarks: first, in view of the
generalization or gauging of the trivial connection $\partial$ in
the flat differential triad $(\A ,\diff,
\partial)$ supporting the (abstract) differential tetrad whose complex is
depicted in (\ref{eq1}) to the non-flat connection $\conn$ in the
`gauged' or `curved triad' $(\A, \diff ,\conn)$, and with our
present sheaf-cohomological interests in mind, we read from
\cite{mall1, mall2} that one can define {\em higher order
cochain-prolongations} $\conn^{i}$ ($i\geq1$) {\em of}
$\conn$($\equiv\conn^{0}$)\footnote{For instance, the first order
prolongation of $\conn^{0}:~\A\mapto\mathbf{\Omega^{1}}$ to
$\conn^{1}:~\mathbf{\Omega^{1}}(\modl)\mapto\mathbf{\Omega^{2}}(\modl)$
is defined by the relation: $\conn(u\otimes v):=u\otimes
dv-v\wedge \conn u$; $u\in\modl(U)$, $v\in\mathbf{\Omega^{1}}(U)$
and $U$ open in $X$. The rest of the $\conn^{i}$s in (\ref{eq21})
are obtained inductively.}, as follows

\begin{equation}\label{eq21}
\begin{array}{c}
\mathbf{\Omega^{0}}(\modl)\stackrel{\conn\equiv\conn^{0}}{\mapto}\mathbf{\Omega^{1}}(\modl)
\stackrel{\conn^{1}}{\mapto}
\mathbf{\Omega^{2}}(\modl)\stackrel{\conn^{2}}{\mapto}\mathbf{\Omega^{3}}(\modl)
\stackrel{\conn^{3}}{\mapto}\cdots\cr
\cdots\stackrel{\conn^{i-1}}{\mapto}\mathbf{\Omega^{i}}(\modl)\stackrel{\conn^{i}}{\mapto}\mathbf{\Omega^{i+1}}(\modl)
\stackrel{\conn^{i+1}}{\mapto}\cdots
\end{array}
\end{equation}

\noindent which, in view of the fact that $\conn$ is non-flat,
with non-zero curvature $\curv$ defined as

\begin{equation}\label{eq22}
\curv(\conn):=\conn^{1}\circ\conn\equiv\conn^{2}\not=
0,\footnote{From this definition of $\curv$ and the definition of
$\conn$ above it follows that $\curv(\conn)$ {\em is an
$\A$-morphism of the $\A$-modules involved ({\it ie}, a structure
sheaf-preserving morphism) and, in particular, a $2$-form} ({\it
ie}, formally: $\curv\in{\rm
Hom}_{\A}(\modl,\Omega^{2}(\modl))=\mathcal{H}om_{\A}(\modl,\Omega^{2}(\modl))(X)$,
so that when $\modl$ is a vector sheaf:
$\mathcal{H}om_{\A}(\modl,\Omega^{2}(\modl))(X)=\Omega^{2}(End\modl)(X)$).
Moreover, we read again from \cite{mall1, mall2}, the curvature
$\curv$ of an $\A$-connection $\conn$ can be viewed as a
$0$-cocycle of local $n\times n$ matrices ($n={\rm dim}\modl$)
having for entries local sections of $\mathbf{\Omega^{2}}$ ({\it
ie}, local $2$-forms on $X$). We will return to this remark in
section \ref{sec7}.}
\end{equation}

\noindent {\em are non-exact}\footnote{We bring the reader's
attention to the fact that in (\ref{eq22}):
$\modl=\A\otimes_{\A}\modl\equiv\mathbf{\Omega^{0}}(\modl)$
\cite{mall1, mall2}. Also, interestingly enough, this definition
of curvature, and in a discrete context similar to ours, was given
in \cite{dim2} (see also \cite{malrap}); moreover, this very
definition for the curvature of an $\A$-connection was used by
Connes in his NDG \cite{kastle, connes}.}. Thus, the obstruction
of the $\conn$-cochain in (\ref{eq21}) to comprise an exact
complex is essentially encoded in the non-vanishing curvature
$\curv$ of the connection $\conn$\footnote{For example,
section-wise in the relevant sheaves:
$(\conn^{i+1}\circ\conn^{i})(s\otimes t)=t\wedge\curv(s)$, with
$s\in\Gamma(U,\modl)$, $t\in\Gamma(U,\mathbf{\Omega^{i}})$ and $U$
open in $X$.}. Thus, $\curv(\conn)$ represents not only the
measure of the departure from differentiating flatly, but also the
deviation from setting up an (exact) cohomology based on
$\conn$---altogether, a measure of the departure of $\conn$ from
nilpotence.

Second, we note in connection with the aforementioned
`section-wise' spirit in which ADG is developed in \cite{mall1,
mall2} that the usual $\smooth$-smooth $\A$-connection---the
connection of a $\smooth$-smooth manifold $X$ the points of which
are coordinatized and individuated by the coordinate algebras in
the structure sheaf $\A=\mathbf{{}^{\com}\smooth}(X)$---acts (as a
$1$-form) on the sheaf of germs of sections of the (complexified)
tangent bundle of $X$:
$T_{\mathbf{C}}(X):=T(X)\otimes_{\R}\mathbf{C}$, the latter
sections being, of course, complex vector fields ({\it ie}, first
rank contravariant tensors over, that is to say, with coordinates
in, ${}^{\com}\smooth(X)$) \cite{manin, mall1, mall2, malrap}. As
a matter of fact, the coordinate structure sheaf
$\mathbf{{}^{\com}\smooth}(X)$ of the $\smooth$-manifold $X$ is
{\em fine}; moreover, every $\mathbf{{}^{\com}\smooth}(X)$-module
sheaf $\modl$ (or its differential counterpart $\diff$) over it is
also fine, hence {\em acyclic}. This is reflected in the well
known `existence result' that {\em every $\smooth$-smooth
$\mathbf{C}$-vector bundle ({\rm equivalently},
$\mathbf{C}$-vector sheaf) on $X$ admits a $\smooth$-connection}
\cite{mall1, mall2}.

The third remark concerns a fundamental difference between a
connection $\conn$ on a vector sheaf and its associated curvature
$\curv(\conn)$, which difference, in turn, bears on a
significantly different physical interpretation that these two
objects have in our finitary theory in particular and more
generally in ADG\footnote{This difference of interpretation
between $\conn$ and $\curv$ will come in handy subsequently when
we discuss and wish to interpret the Chern-Weil theorem.}. We note
that, while according to the definitions given above $\curv$
respects or preserves the abelian algebra structure (or
$c$-coefficient) sheaf $\A$, $\conn$ does not ({\it eg}, it obeys
Leibniz's rule). Since in our scheme $\A$ represents {\em our}
(local) measurements or coordinatizations relative to a (local)
coordinate gauge $\gauge$ that {\em we} lay out to cover and
measure the events of whatever virtual geometric base `space' $X$
{\em we} suppose to be `out there' suitable or convenient enough
for soldering or localizing our algebraic structures, $\curv$ is a
geometric object with respect to these measurements or
`coordinatization actions' in $\A$ ({\it viz.}, $\A$ essentially
encodes the geometry of the background space $X$ \cite{mall1,
mall2})---a kind of `$\A$-tensor', while $\conn$ cannot qualify as
such\footnote{This is in line with what we said earlier about
$\conn$, namely, that it is essentially of algebraic, not
geometric, character. A similar tensor/non-tensor distinction is
familiar to physicists that, as we noted earlier, tend to identify
connection with the gauge potential part $\aconn$ of $\conn$,
since, as it is well known, $\aconn$ transforms non-tensorially
({\it ie}, inhomogeneously) under a gauge transformation, while
$\curv$ obeys a homogeneous, tensorial gauge transformation law.}.
All in all, $\curv$ ({\it ie}, field strength) is what we
measure---a geometric object with respect to our local
measurements/gauge coordinates in $\A(X|_{\gauge})$---when $\conn$
effectively eludes them as well as the background space $X$
supporting them ({\it ie}, serving as a base space for the
structure sheaf $\A$).

In the same train of thought, and following the
(fin)sheaf-theoretic formulation of the principle of (general)
covariance in \cite{malrap} which holds that the laws of nature
are equations between appropriate sheaf morphisms (the main sheaf
morphisms involved being the connection and, more importantly, its
curvature, which, in turn, implies that the laws of Nature are
differential equations, as commonly intuited), we infer that {\em
the laws of physics are independent of our own measurements in
$\A$}, or equivalently, that {\em they are
$\A$-covariant}\footnote{In particular, for the
(fin)sheaf-theoretic expression of the law of gravity in the
absence of matter ({\it ie}, the so-called vacuum Einstein
equations): $\curv_{\rm Ricci}=0$ \cite{mall3, mall4, rap6}, the
aforesaid $\A$-covariance of $\curv_{\rm Ricci}$ indicates the
independence of the law of gravity from our measurements (with
respect to the local gauges in $\gauge$ that we have laid out to
chart $X$) and, ultimately, from the geometry of the background
space $X$ as the latter is encoded in the structure sheaf $\A$
\cite{mall1, mall2, malrap}.}. Also in this line of thought, we
may re-raise the second question opening this paper in another
manner: is it really right to say that the laws of physics ({\it
eg}, gravity) breakdown at singularities if the latter are
diseases that assail our own coordinate algebra sheaves $\A$,
especially when the very mathematical expression of these laws are
independent of (or covariant with) these $\A$s? Stated in a
positive way: the laws of Physis cannot conceivably depend on our
contingent measurements ({\it viz.}, `geometries' and `spaces', or
$\mathbf{C}$-algebraized spaces $(X,\A)$), which in turn means
that when a dynamical law appears to be singular or anomalous
relative to a particular choice of `space-geometry' $(X,\A)$, the
problem does not lie with the law {\it per se}, but, more likely,
with the $\mathbf{C}$-algebraized space that we have
assumed\footnote{For instance, the singularities that assail
general relativity---the classical theory of gravity---are most
likely due to the assumption of coordinate algebras of infinitely
differentiable functions $\A\equiv\smooth(X)$ on a
$\smooth$-smooth spacetime manifold $X$, and are not the `fault'
of Einstein's equations (and the differential mechanism supporting
them) whatsoever.}. Presumably, by changing theory, ultimately,
(modes or operations of) observation and (algebras of)
measurements modelling the latter\footnote{In Greek, the words
`theory' (`$\theta\epsilon\omega\rho\iota\alpha$'), `observation'
(`$\pi\alpha\rho\alpha\tau\eta\rho\eta\sigma\iota\varsigma$') and
`measurement' (`$\mu\epsilon\tau\rho\eta\sigma\iota\varsigma$') go
hand in hand.} (in our scheme, by changing $\A$ and the base space
$X$ supporting this geometry), the apparent singularities can be
resolved, for what could it possibly mean, for instance, if one
could write down Einstein's equations (as in \cite{mall3, mall4})
over ultra-singular (from the $\smooth$-smooth manifold viewpoint)
spaces (as in \cite{malros1, malros2}) other than that the law of
gravity (and the differential apparatus supporting it) does not
depend on the geometry of the background space(time)? We return to
this caustic point in the concluding section.

So, finally, following \cite{malrap}, we are now in a position to
apply ADG's definition of $\A$-connection and define non-trivial
finitary (Lorentzian)
$\A$($\equiv\mathbf{\Omega^{0}_{m}}$)-connections, in complete
analogy with (\ref{eq20}), as

\begin{equation}\label{eq23}
\conn_{m}:=\partial_{m}+\aconn_{m}
\end{equation}

\noindent on the curved (principal) finsheaves of incidence
algebras\footnote{In \cite{malrap}, the structure group of these
$\mathcal{G}$-sheaves was seen to be a finitary version of the
local (orthochronous) spin-Lorentz Lie group of general
relativity; hence, the epithet `Lorentzian' to the
($sl(2,\com)_{m}\simeq so(1,3)^{\uparrow}$-valued) $\aconn_{m}$s
above.} in their corresponding finitary differential triad
$\mathcal{T}_{m}:=(P_{m},\mathbf{\Omega_{m}},\bkd_{m})$\footnote{The
procedure that leads to (\ref{eq23}) was coined `symmetry
localization' or `gauging' in \cite{malrap}, so perhaps one could
also call the corresponding triads `gauged fintriads'
$\mathcal{T}_{m}^{g}:=(P_{m},\mathbf{\Omega_{m}},\conn_{m}=\bkd_{m}+\aconn_{m})$.}
The associated non-zero finitary curvature is denoted by
$\curv_{m}(\conn_{m})$.

\subsection{The abstract WIT, CWT and their finitary analogues}

To make our way towards sheaf-cohomologically classifying the
spin-Lorentzian $\aconn_{m}$s, we first define a de Rham $p$-space
{\it \`a la} \cite{mall1, mall2}. This is just a paracompact
Hausdorff base space $X$ together with an exact de Rham complex as
in (\ref{eq1}) such that the latter's cochain sequence ends at
some grade $p\in\N$, as follows

\begin{equation}\label{eq24}
\cdots\longrightarrow\mathbf{\Omega^{p}}\stackrel{d^{p}\equiv
d}{\longrightarrow}d\mathbf{\Omega^{p}}\longrightarrow 0
\end{equation}

\noindent Then, an important lemma for the (abstract)
WIT\footnote{Following \cite{mall1}, we may coin this lemma `the
generalized Weil Integrality theorem' for reasons to become clear
shortly. Its connection with the usual WIT was first conceived in
\cite{mall}.} states that

\begin{quotation}

\noindent{\em given such an abstract de Rham $p$-space, with every
closed $p$-form $\omega$}\footnote{The reader may recall that
$\omega\in\Omega^{p}$ is said to be closed when $d\omega=0$.} {\em
there is associated a $p$-dimensional \v{C}ech cohomology class
$c$ of $X$ with constant complex coefficients, that is,
$c(\omega)\in\check{H}^{p}(X,\mathbf{C})$}\footnote{Even more
generally, one could replace the constant coefficient sheaf
$\mathbf{C}$ by the $\mathbf{C}$-vector space sheaf ${\rm
ker}\partial$ to arrive to the generalized WIT also employed in
ADG.}.

\end{quotation}

The WIT is a particular consequence of the general lemma above by
taking $p=2$, and it states that

\begin{quotation}

\noindent{\em every $2$-dimensional integral cohomology class
arises as the characteristic class of the curvature $\curv$ of an
$\A$-connection on a line sheaf $\mathcal{L}$}; while, conversely,
that {\em $\curv(\conn)|_{\mathcal{L}}$ yields a (\v{C}ech)
cohomology class in $\check{H}^{2}(X,\Z)$}\footnote{The connection
is clear between this expression of the WIT and the generalized
lemma above; in particular, the integer coefficients arise from
the canonical embedding of the constant sheaf of integers
$\mathbf{Z}$ to the constant sheaf of complexes $\mathbf{C}$ ({\it
ie}, $\mathbf{Z}\stackrel{\subset}{\longrightarrow}\mathbf{C}$)
which, in turn, gives rise to an analogous morphism between the
respective $2$-dimensional sheaf-cohomology groups:
$H^{2}(X,\mathbf{Z})\longrightarrow H^{2}(X,\mathbf{C})$.}.

\end{quotation}

Closely related to the general and concrete WITs above, and lying
at the heart of the theory of characteristic classes, is the CWT
which states that

\begin{quotation}

\noindent{\em given a de Rham $q$-space (with $q$ even) and a
vector sheaf $\mathcal{V}$ of rank $n$ on $X$ endowed with an
$\A$-connection $\conn$ whose curvature is $\curv$}, {\em if $p$
is an invariant polynomial in
$\mathbf{C}[\lambda_{\alpha\beta}]\,\, (1\leq \alpha ,\beta\leq
n)$ of degree $q/2$, then the characteristic closed $q$-form
$\omega$ of the de Rham $q$-space secured by the generalized lemma
for the WIT above can be obtained by identifying
$\lambda_{\alpha\beta}$ with $\curv$\footnote{Where now,
$\lambda_{\alpha\beta}$ are the entries of the $n\times n$-matrix
of (sections of) $2$-forms constituting $\curv(\conn)$ as defined
above \cite{mall1, mall2}.} in $p$ ({\it ie}, from the generalized
WIT in the aforesaid lemma: $c(p(\curv(\conn)))\equiv
c(\curv)\in\check{H}^{q}(X,\mathbf{C})$\footnote{And plainly:
$p(\curv)\in\bigwedge^{q}(\Omega^{1}(X))\stackrel{\subset}{\mapto}
(\bigwedge^{q}\Omega^{1})(X)=\Omega^{q}(X)$.}, {\rm and, more
importantly,} \underline{all this is independently of the given
$\A$-} \underline{connection $\conn$}}\footnote{Which pretty much
vindicates the interpretational distinction that we drew earlier
between the algebraic character of $\conn$ and the geometric
character of its associated curvature $\curv(\conn)$, since the
same `effect' that we measure ({\it viz.}, the geometric object
$\curv$ which is interpreted as the field strength) can in
principle arise from two different `causes' ({\it viz.}, the
algebraic in character $\A$-connection $\conn$ which is
interpreted as the (gauge) potential field). The geometry (and its
supporting space(time)!) that we perceive does not uniquely
determine the algebraic-dynamical substratum (foam) from which it
originates (by our acts of measurement). We are thus tempted,
conceptually at least, to put $\conn$ at the quantum (algebraic)
side, while $\curv$ at the classical (geometrical) side of the
quantum divide, so that an analogue of Bohr's correspondence
principle would be that the classical (commutative) geometric
realm in which $\curv$ lives (together with the $\A$ that it
respects and the $X$ that the latter algebras are supposed to
coordinatize and which essentially supports $\curv$) arises from
measuring ({\it ie}, `observing') the quantum non-commutative
algebraic realm (fluctuating pool, or `quantum foam'
\cite{malros2, rapzap2}) from which $\conn$ derives and in which
it varies.}.

\end{quotation}

The translation of the abstract (vector) sheaf-theoretic versions
of the WIT and CWT above to our finitary case of interest is
immediate: the theorems still hold true in our reticular
environment, because the incidence algebra finsheaves involved
fulfill all the basic technical requirements of ADG for
implementing these theorems vector and algebra sheaf-theoretically
\cite{mall1, mall2, malrap} .

\subsection{En route to classifying the spin-Lorentzian
$\mathbf{\aconn_{m}}$s}

So far one of the main successful applications of ADG is to
sheaf-cohomologically classify Maxwell fields as connections on
line sheaves \cite{sel83, mall1, mall2, mall4}. Here we briefly
expose this application and by analogy we speculate on a possible
finsheaf-cohomological classification of the finitary
spin-Lorentzian connections $\aconn_{m}$ introduced in
\cite{malrap}. Again, we draw information principally from
\cite{mall2}.

\vskip 0.1in

\underline{\bf The Picard Group:} For the cohomological
characterization of vector sheaves, ADG employs sheaf-cohomology;
in particular, it uses their so-called {\em coordinate
$1$-cocycles} to classify them. So, let us dwell for a while on
such a classification scheme.

First, let us assume a $\mathbf{C}$-algebraized space $(X,\A)$ and
a vector sheaf $\modl$ of rank or dimensionality $n$. Let us also
assume an open cover of $X$ or local gauge system for $\modl(X)$,
$\gauge=\{ U_{i}\}|_{i\in I}$, with respect to which one obtains
the following standard Whitney-type of $\A |_{U_{i}}$-isomorphisms

\begin{equation}\label{eq25}
\modl_{i}\equiv\modl|_{U_{i}}\stackrel{\phi_{i}}{\cong}\A^{n}|_{U_{i}}=
(\A|_{U_{i}})^{n}\equiv\A^{n}_{i},\, i\in I\footnote{Where,
$\{\phi_{i}\}$ is a strictly positive partition of unity of $\A$
relative to $\gauge$: $\{\phi_{i}\}\subseteq{\rm End}\A=\A(X)$.}
\end{equation}

\noindent Thus, for any pair of non-trivially intersecting local
open gauges $U_{i}$ and $U_{j}$ in $\gauge$ ({\it ie}, $U_{i}\cap
U_{j}\not=\emptyset$), one obtains the following `local coordinate
change' $\A|_{U_{ij}}$-isomorphism

\begin{equation}\label{eq26}
\phi_{ij}\equiv\phi_{i}\circ\phi_{j}^{-1}\in\mathcal{A}ut_{\A|_{U_{ij}}}
(\A^{n}|_{U_{ij}})={\rm
GL}(n,\A(U_{ij}))=\mathcal{GL}(n,\A)(U_{ij})\footnote{Where the
calligraphic `$GL$', `$\mathcal{GL}$', denotes the $n$-dimensional
general $\A$-linear (structure) group sheaf over $X$.}
\end{equation}

\noindent In fact, such a family of local automorphisms of the
$\A$-module $\A^{n}$ provides a $1$-cocycle of $\gauge$ with
coefficients in the structure group sheaf $\mathcal{GL}(n,\A)$ in
view of the relation

\begin{equation}\label{eq27}
\phi_{ik}=\phi_{ij}\circ\phi_{jk},
\end{equation}

\noindent with $U_{ijk}\equiv U_{i}\cap U_{j}\cap U_{k}$
($i,j,k\in I$). So, we have

\begin{equation}\label{eq28}
\phi_{ij}\in Z^{1}(\gauge ,\mathcal{GL}(n ,\A))
\end{equation}

\noindent which is coined the {\em coordinate $1$-cocycle of
$\modl$ associated with the given local coordinate gauge $\gauge$
of $\modl(X)$}.

So, given that the first homology group of $X$ with coefficients
in $\mathcal{GL}(n,\A)$ is (by definition) the direct limit of the
corresponding \v{C}ech first cohomology group of $X$ as the local
frame $\gauge$ ranges over all covers of $X$, symbolically,

\begin{equation}\label{eq29}
H^{1}(X,\mathcal{GL}(n,\A))=\underrightarrow{\lim}_{\gauge}H^{1}(\gauge
,\mathcal{GL}(n,\A))\footnote{That the (locally finite) open
covers of $X$ form a direct or inductive system was explained in
\cite{rap2}.},
\end{equation}

\noindent one infers that the elements of $H^{1}(\gauge
,\mathcal{GL}(n,\A))$ are equivalence classes of coordinate
$1$-cocycles of $n$-dimensional vector sheaves (denoted as
$[\phi_{ij}]$).

Thus, the sheaf-cohomological classification scheme for vector
sheaves of rank $n$ reads:

\begin{quotation}

\noindent{\em Any $n$-dimensional vector sheaf $\modl$ on $X$ is
uniquely determined by a coordinate $1$-cocycle in $Z^{1}(\gauge
,\mathcal{GL}(n,\A))$ associated with any local gauge $\gauge$ of
$\modl(X)$}. We write $\mathbf{\Phi}^{n}_{\A}(X)$ for the
equivalence (isomorphism) classes of vector sheaves of rank $n$
({\it ie},
$\mathbf{H}^{1}(X,\mathcal{GL}(n,\A))=\mathbf{\Phi}^{n}_{\A}(X)$).

\end{quotation}

In keeping with the section-wise spirit in which ADG is developed,
we note that the equivalence relation between two vector sheaves'
classes $[\mathcal{V}_{1}]$ and $[\mathcal{V}_{2}]$ in
$\mathbf{\Phi}^{n}_{\A}(X)$ can be represented as a similarity
between the section-matrices of their corresponding $1$-cocycles
(say, $v^{1}_{ij}$ and $v^{2}_{ij}$) relative to a common local
chart $\gauge$ covering and coordinatizing $X$, as follows

\begin{equation}\label{eq30}
v^{2}_{ij}=c_{i}\circ v^{1}_{ij}\circ c_{j}^{-1}
\end{equation}

\noindent where $c_{i}\in C^{0}(\gauge ,\mathcal{GL}(n,\A))$ (a
$0$-cochain of $X$ relative to $\gauge$) and $U_{i}\cap
U_{j}\not=\emptyset$, as usual.

Now, in order to make direct connection, as we wish to do here,
with the classification of the bosonic connections $\aconn_{m}$
({\it viz.}, the `{\em finitary quantum causal gauge potentials}')
on the curved finsheaves of incidence algebras representing the
kinematics of dynamical quantum causality in \cite{malrap}, we
follow Selesnick's axiomatics for line bundle-classification in
\cite{sel83}, only here we are obviously interested in line
sheaves ({\it ie}, vector sheaves of rank $1$).

So, from \cite{mall1, mall2} we read that for $n=1$ we get the
following isomorphism

\begin{equation}\label{eq31}
\mathbf{\Phi}^{n}_{\A}(X)=H^{1}(X,\A^{^\centerdot})\footnote{Where
$\A^{^\centerdot}$ denotes the sheaf of invertible elements
(units) in the commutative algebra structure sheaf $\A$, which
means that $\A^{^\centerdot}$ is an abelian group sheaf.}
\end{equation}

\noindent something that, without going into too much detail,
enables us to arrive at the so-called {\em Picard group of
$X$}---an abelian group consisting of equivalence classes of line
bundles on $X$---and defined as follows

\begin{equation}\label{eq32}
{\rm
Pic}(X):=(\mathbf{\Phi}^{1}_{\A}(X),\otimes_{\A})\equiv\mathbf{\Phi}^{1}_{\A}(X)
\end{equation}

\noindent where the commutative and associative tensor product
functor $\otimes_{\A}$ has been employed to endow
$\mathbf{\Phi}^{n}_{\A}(X)$ in expression (\ref{eq30}), and for
$n=1$, with an abelian group structure\footnote{Where, it is
understood that the tensor product
$\mathcal{L}\otimes_{\A}\mathcal{L}^{'}$ of two line sheaves is a
line sheaf whose coordinate $1$-cocycle is the
$\otimes_{\A}$-product of the $1$-cocycles of the corresponding
line sheaves (closure with respect to $\otimes_{\A}$-operation),
that the inverse of a line sheaf $\mathcal{L}$ is its dual
$\mathcal{L}^{-1}=\mathcal{L}^{*}=\mathcal{H}om_{\A}(\mathcal{L},\A)$
(inverse), and that the neutral element is the structure sheaf
$\A$ itself, since:
$\mathcal{L}\otimes_{\A}\mathcal{L}^{*}=\mathcal{H}om_{\A}(\mathcal{L},\mathcal{L})
\equiv\mathcal{E}nd\mathcal{L}=\A$ (neutral element).}.

With the Picard group in hand, ADG achieves a sheaf-cohomological
classification of Maxwell connections $\conn_{Max}$ on line
sheaves by making use of the so-called {\em Chern isomorphism}:

\begin{equation}\label{eq33}
H^{1}(X,\A^{^\centerdot})=H^{2}(X,\mathbf{Z})\stackrel{\rm
WIT}{\Longrightarrow}[\curv_{Max}]\in{\rm
im}(H^{2}(X,\mathbf{Z})\mapto H^{2}(X,\mathbf{C}))
\end{equation}

\noindent which is essentially a consequence of (the abstract
version of) CWT as it translates the problem of classifying
$\conn_{Max}$ on line sheaves {\it per se} to one of {\em finding
the equivalence classes of Maxwell fields having a given curvature
$2$-form $\mathcal{F}_{Max}$}\footnote{It is important to mention
at this point that by a {\em Maxwell field} ADG means a pair
$(\mathcal{L},\conn_{Max})$ consisting of a line sheaf
$\mathcal{L}$ and a Maxwellian $\A$-connection $\conn_{Max}$ on
it. $\mathcal{L}$ is interpreted as `{\em the carrier space of
$\conn_{Max}$}'---and only because of the line sheaf carrying it a
connection may be regarded as a geometric entity (but certainly
not transformation-wise, {\it ie}, tensorially speaking).}.
Moreover, in view of ADG's quantum interpretation of connection
$\conn$ and its curvature $\curv(\conn)$ in footnote 98, we may
read the sheaf-cohomological classification of Maxwell fields via
the Chern isomorphism above in a quantal way: what we actually
determine ({\it viz.}, `measure' or `observe') is the `classical',
`commutative' (since it respects the abelian coordinatizations in
$\A$) field strength $\curv$, while the algebraic or quantal
`causes' (or origins) of a given (measured) $\curv$ remain {\em
indeterminate}, since {\em a given $\curv$ corresponds to a whole
cohomology class of connections}! This indeterminacy resembles,
even if only in spirit, Heisenberg's standard one and it accords
with our insistence in footnote 98 on placing the algebraic in
nature $\conn$ on the quantum side of Heisenberg's {\it schnitt},
while its geometric in character $\curv(\conn)$ in the classical
realm on the other side of the quantum divide. For recall (a
watered down version of) Bohr's Correspondence Principle: from the
noncommutative `quantum soup' we always extract ({\it ie},
measure) commutative numbers. In fact, all this agrees with the
very interpretation of the term `spacetime foam' in \cite{malros2}
and its finitary-algebraic in \cite{rapzap2}.

So, the finsheaf-cohomological classification of the non-trivial
({\it ie}, curved) spin-Lorentzian connections $\aconn_{m}$ on the
principal finsheaves of qausets defined in \cite{malrap} follows
directly from the analogous classification scheme of the
$\conn_{Max}$ above, since, as it was repeatedly stressed
throughout the present paper and partly in \cite{malrap}, these
finsheaves fulfill all the requirements of ADG for performing
sheaf-cohomological differential geometric constructions in spite
of the $\smooth$-manifold. Thus, we define a `causon
field'\footnote{In \cite{malrap}, a `causon' was defined to be the
elementary particle of the `reticular bosonic spin-Lorentzian
gauge potential field $\aconn_{m}$ representing local curved
quantum causality', and it was speculated that it must be
intimately related to the graviton---the anticipated quantum of
the gravitational field.} to be the following pair

\begin{equation}\label{eq34}
(\vec{\mathcal{L}}_{Caus}^{m},\vec{\conn}_{Caus}^{m})
\end{equation}

\noindent consisting of a line finsheaf
$\vec{\mathcal{L}}_{Caus}^{m}$ associated to the curved
$\mathcal{G}$-finsheaves $\vec{\mathcal{S}}_{m}$ of qausets in
\cite{malrap}, together with a non-trivial
$\vec{\mathbf{\Omega}}^{\mathbf{0}}_{\mathbf{m}}$-connection
$\vec{\conn}_{Caus}^{m}$ on it\footnote{As explained in
\cite{malrap}, the arrow sign over the relevant symbols above
indicate the (quantum) causal interpretation that these structures
carry. From (\ref{eq34}) it follows that the $\aconn^{m}_{Caus}$
part of $\vec{\conn}^{m}_{Caus}$ should also carry an arrow
(write: $\vec{\aconn}^{m}_{Caus}$) \cite{malrap}. Of course, we
can further remark at this point that we are aware that the photon
(the quantum of $\aconn_{Max}$) is a spin-$1$ gauge boson, while
the graviton, a spin-$2$ quantum. Here, however, we do not intend
to dwell longer on the spin-particulars of the causon
$\vec{\aconn}^{m}_{Caus}$ other than that, quantum
spin-statistically speaking, it is a boson \cite{sel83, mall1,
mall2}.}.

In connection with the above, we still remark that the
aforementioned `{\em Selesnick's Correspondence Principle}'
\cite{sel83, mall4, mall6} is used herewith in the (sheaf)
topological algebra theory setup, when usually referring to the
topological ({\em not} Banach) algebra of smooth functions on a
(compact) manifold, based on a $K$-theory argument, providing
further, directly, the `{\em smooth analogue}' of the classical
(`continuous') Serre-Swan theorem \cite{mall6}---or in more
detail, \cite{mall4}. In keeping with Selesnick's vector bundle
axiomatics in \cite{sel83}, as well as with its vector sheaf
descendants in \cite{mall1, mall2, mall4, mall6}, local sections
of the $\vec{\mathcal{L}}_{Caus}^{m}$s in (\ref{eq34}) correspond
to local (pre)quantum\footnote{The epithet `prequantum' pertains
to a possible application of the general theory of `geometric
prequantization' as developed in \cite{mall1, mall2, mall5, mall4,
mall6} to the causon field in (\ref{eq34}). See subsection 7.1
next.} states of bare or free causons. This brings us to the next
section.

\section{Future outlook: a couple of applications to discrete Lorentzian
quantum gravity}\label{sec7}

In this penultimate section we discuss two possible future
applications of some of the ideas that were put forward above to
certain aspects of current discrete Lorentzian quantum gravity
research that are of interest to us.

\subsection{Geometric prequantization of Lorentzian gravity}

Continuing the remarks that conclude the last section (and
footnote), we note that according to Selesnick's general
$\smooth$-vector bundle axiomatics in \cite{sel83}

\begin{quotation}

\noindent{\em local sections of line bundles correspond to states
of free bosons}; while, {\em local sections of vector bundles
(such as $\mathbf{\Omega^{1}}$) correspond to states of bare
fermions}.

\end{quotation}

\noindent ADG's vector sheaf analogues of these results, as
explained above, are of immediate avail:

\begin{quotation}

\noindent {\em boson states are sections of line sheaves, while
fermion states are sections of Grassmannian (exterior) vector
sheaves}

\end{quotation}

\noindent and, of course, ADG's generality allows us to consider
not only smooth vector sheaves, but {\em any vector sheaf over, in
principle, any base space}\footnote{Which serves as a base
space(time) ({\it viz.}, `configuration space') for the physical
system in focus.}. An important immediate application of the
foregoing ideas, and in particular of WIT, is the result, just
quoted {\it verbatim} from \cite{mall1, mall2, mall5, mall6},
that:

\begin{quotation}
\noindent {\em Every free elementary particle is prequantizable;
that is to say, it entails by itself a prequantizing line
sheaf\footnote{In fact, this is so regardless of whether the
elementary particle is a boson or fermion \cite{sel83, mall,
mall1, mall2, mall5, mall6}.}.}
\end{quotation}

\noindent and in our particular finsheaves of qausets scenario for
discrete Lorentzian gravity \cite{malrap, rap6}, that:

\begin{quotation}

\hskip 1.0in {\em A free causon entails by itself
$\vec{\mathcal{L}}_{Caus}^{m}$}.

\end{quotation}

\noindent In a nutshell, the importance of this result is that, in
line with the general philosophy of geometric quantization
\cite{souriau, simms, wood, mall5, mall6}, one is able to arrive
at the main constructions of quantum field theory ({\it ie},
conventionally speaking, 2nd-quantized structures) by {\em
avoiding altogether the process of 1st-quantization}, thus,
effectively, by {\em avoiding altogether any fundamental
commitment to the classical Hamiltonian mechanics and the
so-called `canonical formalism' that accompanies it}. For the case
of (the quantization of) gravity in particular, such a scheme
\cite{souriau} would appear to bypass in a single leap the whole
of the canonical approach to quantum gravity with all its
technical and conceptual problems. Just to mention three such
problems:

\begin{itemize}

\item {\bf (a)} The problem of the diffeomorphism
group ${\rm Diff}(M)$---the gauge group of general
relativity---since the canonical theory assumes a background
differential ({\it ie}, $\smooth$-smooth) manifold spacetime $M$.

\item {\bf (b)} The problem of finding the
`right' (Hilbert) physical state space $\mathcal{H}$ for the
graviton---with the notorious problems of time, unitarity and
probability interpretation in quantum gravity that go with it.

\item {\bf (c)} The problem of deciding {\it prima facie} ({\it ie},
straight from the classical theory in some rather `natural' way)
what are (the algebras of) the physical observables (to be
represented in $\mathcal{H}$ above) relevant to quantum gravity,
since, for instance, there are quantum mechanical observables
without known classical counterparts \cite{bohm}\footnote{Thus, it
would be begging the question to (canonically) quantize a
classical theory---in particular, general relativity---since we
could encounter entities in the quantum regime that are not
observable at the classical level (in which case, the
correspondence principle would be effectively meaningless).}.

\end{itemize}

\noindent and it is clear from the foregoing how the application
of geometric prequantization {\it \`a la} ADG to a finitary,
causal and quantal version of Lorentzian gravity \cite{malrap,
rap6} may be able to evade all three. At this point we could also
infer that the finsheaf-theoretic scenario for discrete Lorentzian
quantum gravity via ADG is more in line with a covariant path
integral (over spaces of self-dual $sl(2,\com)_{m}$-valued
$\aconn_{m}$s) approach to the quantization of gravity, rather
than with the canonical (Hamiltonian) scheme. This too was
anticipated at the end of \cite{malrap}.

\subsection{Finitary ADG on consistent-histories, topoi in quantum
logic and quantum gravity, and a connection with SDG}

The second future application of the finitary ADG ideas above that
we would like to suggest is to the consistent-histories approach
to quantum theory and quantum gravity in particular.

In \cite{rap5}, for instance, sheaves of qausets over the
Vietoris-topologized base poset category of Boolean subalgebras of
the universal orthoalgebra of history propositions were
introduced, as it were, to define {\em sheaves of quantum causal
histories}. At the end of the paper it was speculated that one
should be able to do differential geometry {\it \`a la} ADG on
such sheaves---something that could be of immediate value to
quantum gravity research when approached via consistent-histories.
There seems to be no foreseeable obstacle to such an endeavor,
since, as we have time and again stressed, the results of ADG are
effectively base space independent\footnote{The reader is
encouraged to read the concluding remarks in \cite{rap5} that
predict, for example, a possible sheaf-cohomological
classification of the algebra sheaves of quantum causal histories
along the lines of ADG. See also the following paragraph.}.

A more specific project along these lines could be the following:
since the topos-theoretic perspective on both the quantal logic of
consistent-histories \cite{isham97} and on the usual quantum logic
\cite{buttish1, buttish2, buttish3} has revealed to us that in a
very geometrical sense quantum logic is warped or curved relative
to its local classical sublogics, so the closely analogous
topos-like aggregate of quantum causal histories'
sheaves\footnote{Coined the `Quantum Causal Histories Topos'
(QCHT) in \cite{rap5}.} may also exemplify such a curvature which,
in view of the quantum causal interpretation of the objects in the
QCHT, may be directly related to the reticular curved quantum
causality ({\it viz.}, discrete Lorentzian quantum gravity)
studied in \cite{malrap}\footnote{That a topos-theoretic approach
not only to quantum logic, but also to quantum gravity proper, is
quite a promising route, was nicely presented in \cite{buttish4}.
See also \cite{rap4}.}. Thus, for example, it would be interesting
to {\em search for a non-trivial characteristic cocycle in the
curved sheaves of quantum causal histories}. The ideas developed
in this paper clearly indicate that this is a legitimate and quite
feasible project\footnote{This project, in the context of the
topos-theoretic approach to quantum logic proper \cite{buttish1,
buttish2, buttish3} and to the similar approach to the logic of
consistent-histories \cite{isham97}, was originally conceived by
John Hamilton and Chris Isham (Chris Isham in private
communication).}.

\medskip

We close this section with two remarks: first, the aforementioned
possible QCHT organization of the sheaves of quantum causal
histories should be compared with the topoi modelling the
mathematical universes in which to carry out the Kock-Lawvere
Synthetic Differential Geometry (SDG) \cite{lav96}. The latter, in
a nutshell, is an extension of the usual, `classical'
$\smooth$-differential geometry by two means: first, by admitting
nilpotent `real numbers', and second, by suitably modifying the
logic underlying the usual Calculus from the Boolean (classical)
one of the topos $\mathbf{Set}$ of classical constant sets (in
which, for instance, the usual $\smooth$-calculus is constructed),
to the Brouwerian (intuitionistic) one of the topos
$\mathbf{Sh}(X)$ of varying sets \cite{rap5} in order to cope with
the first extension\footnote{It is worth mentioning here the
result, due to Dubuc \cite{dubuc1, dubuc2}, that the category of
(finite dimensional) paracompact $\smooth$-smooth manifolds and
diffeomorphisms between them can be faithfully embedded into a
topos, preserving fiber products, open covers, as well as mapping
the usual real line $\R$ into the aforementioned
nilpotent-enriched `real numbers'---the so-called Kock-Lawvere
ring $R_{KL}$.}. Moreover, SDG purports to be able to translate
almost all the basic constructions of the usual Calculus on smooth
manifolds into synthetic terms\footnote{For instance, it has
provided, like we have done here for the finitary case, a
synthetic version of de Rham's theorem at the level of chain
complexes, and much more...}. Only for this, and in view of
similar claims made about ADG in the present paper, it would
certainly be worthwhile to initiate a comparison between ADG and
SDG, even if only at an abstract mathematical level\footnote{For
instance, it would be interesting to  compare the way the two
theories extend and generalize the usual de Rham theory on
$\smooth$-smooth manifolds.}. However, as far as applications to
quantum gravity are concerned, such a comparison could prove to be
beneficial to physics too, since it has been seriously proposed
that SDG could cast light on the problem of quantum gravity
\cite{buttish4}.

Finally, Finkelstein \cite{df88}, in a reticular-algebraic model
for the quantum structure and dynamics of spacetime similar to
ours, called the `causal net', urged us to develop a causal
version of the (co)homology theory of the usual algebraic
topology---as it were, `{\em to algebraize and causalize (with
ultimate aim to quantize) topology (in order to apply it to the
quantum structure and dynamics of spacetime)}'. Since, following
Sorkin's insight in \cite{sork95} to change physical
interpretation of the fintoposets involved from topological to
causal, our finitary incidence algebras model qausets (and not
topological spaces proper) \cite{rap1} while their curved
finsheaves represent the qausets' dynamical variations
\cite{malrap}, it is perhaps fair to say that the finitary
\v{C}ech-de Rham finsheaf-cohomology presented in the present
paper comes very close to materializing Finkelstein's imperative
above\footnote{Thus, the reader can now go back to the various
(co)homological structures mentioned in the present paper and draw
an arrow (indicating causal, not topological proper,
interpretation) over their symbols!}.

\section{Physico-philosophical finale}\label{sec8}

We close the present paper by making some physico-philosophical
remarks in the spirit of the two questions raised in the
introductory section.

\medskip

We hope that by this work we have made it clear that one can
actually carry out most of the usual differential geometric
constructions effectively without use of any sort of
$\smooth$-smoothness or any of the conventional `classical'
Calculus that goes with it. This to a great extent indicates, in
partial response to the second question opening the paper, that
the differential geometric technique or `mechanism'---the
`differential mechanics' or `differential operationalism' so to
speak---is not crucially dependent on a $\smooth$-smooth
background space(time) and the coordinate algebras of
$\smooth$-smooth functions (or generalized `position
measurements', or even `localizations') associated with its
geometric points, no matter how strongly the usual calculus on
manifolds has `forced' us so far to postulate it up-front before
we set up any differential geometric theory/model of Nature. To
this in many ways misleading pseudo-imperativeness we tend to
attribute the almost instinctive reaction of the modern
mathematical physicist to regard the smooth continuum as a model
of spacetime of great {\em physical} significance and
import\footnote{This brings to mind Einstein's famous suspicion
about the actual physicality of spacetime (``{\it Space and time
are concepts by which we think, not conditions in which we live}''
\cite{einst50}), and its $\smooth$-smooth manifold model (see the
three quotations of Einstein in the opening section).}.
Admittedly, the manifold has served us well; after all, the very
differential geometry on which Einstein's successful general
relativity theory of the (classical) gravitional field rests is
vitally dependent on it.

However, it soon became clear by means of the celebrated
singularity theorems \cite{pen1} that the classical theory of
gravity and the smooth spacetime continuum that supports it are
assailed by anomalies and diseases in the form of singularities
long before a possible quantization scheme for them becomes an
issue. Especially the so-called black hole singularities seem to
indicate that general relativity and its classical continuous
spacetime backbone break down near, let alone in the interior of,
them\footnote{Furthermore, this came to be distilled to the
following Popperian `falsifiability'-like motto: general
relativity is a good theory, because, among other things such as
agreement with experiments/observations, it predicts its own
downfall by the existence of singular solutions to Einstein's
gravitational field equations.}. It now appears plain to us that
classical differential geometry cannot cope with such pathologies
and this has prompted theoretical physicists to speculate that a
quantum theory of gravity should be able to heal or at least
alleviate these maladies. Indeed, Hawking's semi-classical (or
semi-quantum!) treatment of these objects showed us that they
should properly be regarded not as universal absorbers, but as
some kind of thermodynamically unstable black bodies that can
thermally radiate quanta \cite{hawk1}. An even more startling
behavior of such singularities was discovered a bit earlier by
Bekenstein and Hawking \cite{bek, hawk2} who showed that they have
rich thermodynamic and, {\it in extenso}, information-theoretic
attributes not describable, let alone explainable, by a classical
theoresis of spacetime structure and its dynamics. It now seems
natural to the theorist to anticipate that only a cogent quantum
theory of gravity can deal effectively with black hole
physics---especially with their aforementioned thermal evaporation
phenomena and their horizons' area-proportional entropy.

On the other hand, one could also view gravitational singularities
from a slightly different perspective. Such a perspective was
adopted by Finkelstein \cite{df58} when he dealt with the doubly
singular Schwarzschild solution to Einstein's equations. In that
paper he effectively showed, by employing a novel spacetime
coordinate system, that the external singularity ($r=2m$) in the
`epidermis' of the Schwarzschild black hole indicates, in fact,
that the latter is a unidirectional membrane allowing the
propagation through its horizon of particles, but not of their
antiparticles\footnote{Pointing thus to a fundamental
time-asymmetry even in the classical gravitational deep
\cite{df88}.}. At the same time, however, his work also implicitly
entailed that the interior singularity ($r=0$) cannot be done away
with simply by a coordinate transformation, thus indicating that
in the `guts' of the Schwarzschild black hole---right at the
point-mass source of the gravitational field---there is a `real'
singularity ({\it ie}, not just a coordinate one) which signals
the inadequacy of general relativity in describing the
gravitational field right at its source. Again, it is currently
believed that only a quantum theoresis of gravity can achieve such
a description\footnote{This may be understood in close analogy
with QED which effectively gave, with the aid of some
theoretically rather {\it ad hoc} and conceptually questionable
renormalization procedures, a calculationally finite theory of the
interaction of the photon radiation field with its source point
electron. Alas, quantum gravity, when regarded as the quantum
field theory of $g_{\mu\nu}$, like QED is for $A_{\mu}$, can be
shown to be non-renormalizable...}.

To us, what is very educational from Finkelstein's alternative
perspective on the singularity riddle is the employment of new
coordinates (albeit, still labelling the point events of a
classical differential manifold model for spacetime) which
effectively resolved the exterior singularity, followed by a sound
physical interpretation (particle/antiparticle or
past/future-asymmetry) of the resolved picture. This is in
striking contrast to the usual treatment of singularities as real
physical diseases that cannot be cured within the classical
$\smooth$-smooth differential geometric framework of general
relativity \cite{pen1}. Thus, in such `$\smooth$-conservative'
approaches, singularities are not to be encountered, because one
does not know how to treat them: rather, they are to be isolated
and cut-off from a remaining `effective spacetime manifold' in
which non-anomalous physical processes occur normally and can be
adequately described by $\smooth$-smooth means \cite{geroch,
geroch1, geroch2}.

In contradistinction, what we advocate herein is akin, at least in
spirit and philosophy, to Finkelstein's approach: by changing
focus from the classical coordinate structure algebra sheaf of
$\smooth$-smooth functions on the differential manifold to another
structure $\A$-sheaf more suitable to the physical problem under
theoretical scrutiny, while still retaining at our disposal most
of the panoply of the powerful differential mechanism of the usual
$\smooth$-calculus, we effectively integrate, absorb or `engulf'
singularities in our theory rather than stumble onto them and, as
a result, meticulously try to avoid them\footnote{As it were, by
making sure that we avoid them so that we can continue performing
the usual $\smooth$-calculus. In this sense our theory is not
`$\smooth$-calculus conserving'.}. Thus, altogether there is no
issue of avoiding singularities or of continuing to perform
$\smooth$-calculus in a singularity-amputated smooth spacetime
manifold, since we can calculate ({\it ie}, actually carry out an
abstract and quite universal calculus) in their very presence.
Singularities are not impediments to ADG, for its abstract,
algebraic in nature, sheaf-theoretic differential mechanism in a
strong sense `sees through them', while on the other hand, the
classical $\smooth$-differential geometry on smooth manifolds is
quite impervious to and intolerant of them. What this contrast
entails, of course, is that on the one hand mathematicians
(especially differential equations specialists) should tell us
what it means to set up a perfectly legitimate differential
equation with possibly ultra-singular coefficient functions and
look for its solutions within the ultra-singular structure
$\A$-sheaf\footnote{Like in \cite{malros1}, for instance, where
$\A$-sheaves of functions with everywhere dense singularities were
studied under the prism of ADG, or even more generally,
subsequently in \cite{malros2}, where in the same spirit
`multi-foam algebras' dealing with singularities on arbitrary sets
(under the proviso that their complements are dense) were
considered.}, and on the other the theoretical physicist
(especially the relativist and the quantum gravity researcher) is
burdened with the responsibility to {\em physically interpret} `a
dynamics amidst singularities, and in spite of them' much in the
same way that, as briefly noted above, Finkelstein in \cite{df58}
physically interpreted the new picture of the exterior
Schwarzschild singularity in the light of new coordinates as a
semi-permeable ({\it ie}, particle-allowing/antiparticle-excluding
or equivalently past/future-asymmetric) membrane\footnote{In a
`psychological' sense, one is expected to be surprised or even
intimidated, hence one's calculus to be impeded, by singularities
when one works in the featureless and uniform differential
manifold and its $\smooth$-algebras of coordinates; while on the
other hand, if singularities is what one routinely encounters in
the space and its coordinate functions that one is working with
like, for instance, in \cite{malros1, malros2}, and at the same
time one is able to retain most of the practically useful
differential mechanism, one is hardly in awe of singularities, so
that one proceeds uninhibited with one's differential geometric
constructions and singularities present no essential problem.}.

What will certainly burden us in the immediate future is to set up
a finitary version of Einstein's equations in the language of ADG,
since they have already been cast abstractly in \cite{mall3,
mall4}. Indeed, the second author is already looking into this
possibility \cite{rap6}; furthermore, it must be noted that the
algebraic ideas propounded above are in close analogy with Regge's
famous coordinate-free and reticular simplicial gravity proposed
in \cite{regge} and further elaborated from a (topo)logical
perspective in \cite{zap3}, although it must also be admitted that
the `freedom from coordinates' espoused by ADG is, in fact,
freedom to use in principle {\em any} coordinate algebra structure
sheaf no matter how singular or anomalous it may seem to be from
the conventional $\smooth$-smooth perspective. For how can the
laws of Nature, that are usually described in terms of
differential equations, stumble upon our own measurements, on our
own coordinatizations ({\it ie}, `arithmetizations' or
`geometrizations') of Her events (phenomena) and the spaces that
host them? How can we ever hope to understand Physis if we ascribe
to Her singularities and pathologies when it is more likely that
it is our own theories that are short-sighted, of limited scope
and descriptive power?

It seems only proper to us to conclude the present study as we
started it in section \ref{sec1}, namely, by quoting and briefly
commenting on Einstein, as well as by summarizing, by means of
`sloganizing', our basic thesis:

\begin{quotation}

``...It does not seem reasonable to introduce into a continuum
theory points (or lines {\it etc}) for which the field equations
do not hold... Is it conceivable that a field theory permits one
to understand the atomistic and quantum structure of reality?
Almost everybody will answer with `no' and...at the present time
nobody knows anything reliable about it...so that we cannot judge
in what manner and how strongly the exclusion of singularities
reduces the manifold of solutions... {\em We do not possess any
method at all to derive systematically solutions that are free of
singularities}\footnote{Our emphasis.}.'' (1956) \cite{einst56}

\end{quotation}

We do sincerely hope that, at least conceptually, the ideas
propounded herein will help us catch initial, but nevertheless
clear, glimpses of such an apparently much needed mathematical
method.

Finally, the following two `slogans' crystallize our central
thesis in the present paper:

\vskip 0.1in

\noindent In the same way that

\begin{quotation}

\noindent {\bf Slogan 1.} {\em Continuity is independent of the
continuum.}\footnote{That is, when spacetime is modelled after a
topological ($\cont$) manifold \cite{sork91, rap2}.}

\end{quotation}

\noindent so

\begin{quotation}

\noindent {\bf Slogan 2.} {\em Differentiability is independent of
smoothness.}\footnote{That is, when spacetime is modelled after a
smooth ($\smooth$) manifold \cite{mall1, rapzap1, malrap,
rapzap2}.}

\end{quotation}

\section*{Acknowledgments}

Both authors are indebted to Chris Isham, true appreciator of
`unorthodox' thought and proven supporter of `adventurous'
research, for numerous encouraging exchanges. They would also like
to thank Keith Bowden for useful comments on early drafts of this
paper. The second author also acknowledges long time fruitful
discussions on the possibility of an entirely algebraic
description of causality and how it may shed light on the problem
of constructing a cogent quantum theory of spacetime structure and
its dynamics (gravity) with Steve Selesnick and Roman Zapatrin.
The incessant moral support of Jim Lambek over many years of
research into sheaf and topos theory, and their potential
application to theoretical physics, is also greatly appreciated.
The second author is also grateful to the European Commission for
the material support in the form of a generous Marie Curie
Individual Research Fellowship held at Imperial College of
Science, Technology and Medicine, London (UK).

\end{document}